\documentclass[reprint,
superscriptaddress,
 amsmath,amssymb,
 aps,
pra,
twocolumn
]{revtex4-2}

\usepackage{graphicx}% Include figure files
\usepackage{dcolumn}% Align table columns on decimal point
\usepackage{bbm}% bold math

\usepackage{dsfont}% special shape latters
\usepackage{braket}% Dirac notation
\usepackage{hyperref}% add hypertext capabilities
\usepackage{xcolor}
\usepackage{soul}
\usepackage{ulem}
\usepackage{float}
\usepackage{amsmath}
\usepackage{footnote}
\newcommand{\RNum}[1]{\uppercase\expandafter{\romannumeral #1\relax}}% for Roman number
\usepackage{physics}

\begin{document}

%\preprint{APS/123-QED}

\title{Two-photon correlations and HOM visibility from an imperfect single-photon source}
\author{Eva M. González-Ruiz$^\dagger$}
\email{eva.gonzalezruiz@ipht.fr}
\affiliation{ Center for Hybrid Quantum Networks (Hy-Q), Niels Bohr Institute\\
 University of Copenhagen, Blegdamsvej 17, DK-2100 Copenhagen, Denmark }
 \affiliation{Institut de Physique Théorique, Université Paris-Saclay, CEA, CNRS, 91191 Gif-sur-Yvette, France.
}
\author{Johannes Bjerlin$^\dagger$}
\affiliation{ Center for Hybrid Quantum Networks (Hy-Q), Niels Bohr Institute\\
 University of Copenhagen, Blegdamsvej 17, DK-2100 Copenhagen, Denmark }
\affiliation{Department of Physics and Astronomy, University of Southern California, Los Angeles, California 90089-0484, USA}
\affiliation{Department of Physics, Division of Mathematical Physics, Lund University, Professorsgatan 1, 22363
Lund, Sweden.}
\author{Oliver August Dall'Alba Sandberg}
\affiliation{ Center for Hybrid Quantum Networks (Hy-Q), Niels Bohr Institute\\
 University of Copenhagen, Blegdamsvej 17, DK-2100 Copenhagen, Denmark }
\author{Anders S. S\o rensen}
\affiliation{ Center for Hybrid Quantum Networks (Hy-Q), Niels Bohr Institute\\
 University of Copenhagen, Blegdamsvej 17, DK-2100 Copenhagen, Denmark }

\date{\today}

%******** abstract ********
\begin{abstract}
We study the single photon purity of a resonantly driven single-photon source in the realistic scenario where the excitation laser can leak into the detection path. We find that the duration of the excitation pulse strongly influences the quality of the single-photon source. We calculate the influence of this on the effective parameters describing the most relevant properties, including the two-photon component and Hong-Ou-Mandel (HOM) visibility. Furthermore, we analyze how these properties can be strongly affected by frequency filtering of the outgoing field. 
Our results highlight that the relation between the two-photon component of the emission and the HOM visibility is more complicated than typically assumed in the literature, and depends on the specific details of the source. 

\end{abstract} 
\maketitle

\section{\label{sec-1Intro} Introduction}

One of the crucial components in the development of scalable photonic quantum technologies is reliable single-photon sources~\cite{thomas2021race}. They are necessary both in the implementation of photon-based optical quantum computing~\cite{knill2001scheme,o2007optical}, quantum simulation~\cite{aspuru2012photonic}, and long-distance quantum communication~\cite{duan2001long,sangouard2011quantum}, with the latter point forming the bedrock of the quantum internet~\cite{kimble2008quantum}. 
One avenue towards reliable single-photon sources are single quantum emitters, like semiconductor quantum dots~\cite{tomm_bright_2021, aharonovich2016solid,senellart2017high,wang2016near,arakawa_progress_2020, ollivier_reproducibility_2020, laferriere_unity_2022, lu_quantumdot_2021, valeri_generation_2024}.
Recent advances have included scalable on-demand sources with strings of $>100$ indistinguishable photons~\cite{uppu2020scalable}. Such setups serve as likely candidates for loophole-free violation of Bell's inequality~\cite{gonzalez2022violation} and device-independent QKD \cite{gonzalez2022diqkd}, as well as multiple other applications for the development of a quantum internet \cite{lu_quantumdot_2021} and fault-tolerant quantum computing \cite{maring_versatile_2024}. 

Unfortunately, the often challenging experimental realization of single-photons sources deviate from generating identical and pure single photons on demand. To improve on these properties, it is essential to have a thorough understanding of the imperfections of the sources. The most relevant errors in realistic single-photon sources are typically characterized by the second-order correlation function $g^{(2)}$, describing the multi-photon emission and typically referred to as the impurity of the source, and the Hong-Ou-Mandel (HOM) visibility~\cite{hong1987measurement}, characterizing the indistinguishability of the emitted photons. 

In this article, we study an emitter based, realistic single-photon source driven resonantly by a Gaussian-pulsed coherent field. In realistic settings, this field can leak into the detection mode, which affects the quality of the outgoing state. Our analysis includes frequency filtering of the emission, which is commonly used experimentally to reduce the leakage of the driving light. 
Additionally, we study the relation between the HOM
visibility~\cite{hong1987measurement} and the second order correlation function $g^{(2)}$. Although this relation has been traditionally oversimplified to $V = 1 - 2g^{(2)}$~\cite{somaschi2016near,wang2019towards}, recent work has dug into the problem and investigated how the indistinguishability of photons \cite{ollivier2021hong}, as well as photon-number coherence generation \cite{loredo_generation_2019,wenniger_quantum_2024}, affects the relation between multi-photon component and HOM visibility.
This relation is of particular interest since HOM visibility is typically performed to attain information on the intrinsic indistinguishability of the photons in experiments~\cite{trivedi2020generation}. This can only be achieved if the influence of the multi-photon component is known. 

\begin{figure*}[]
 \includegraphics[width=0.7\textwidth]{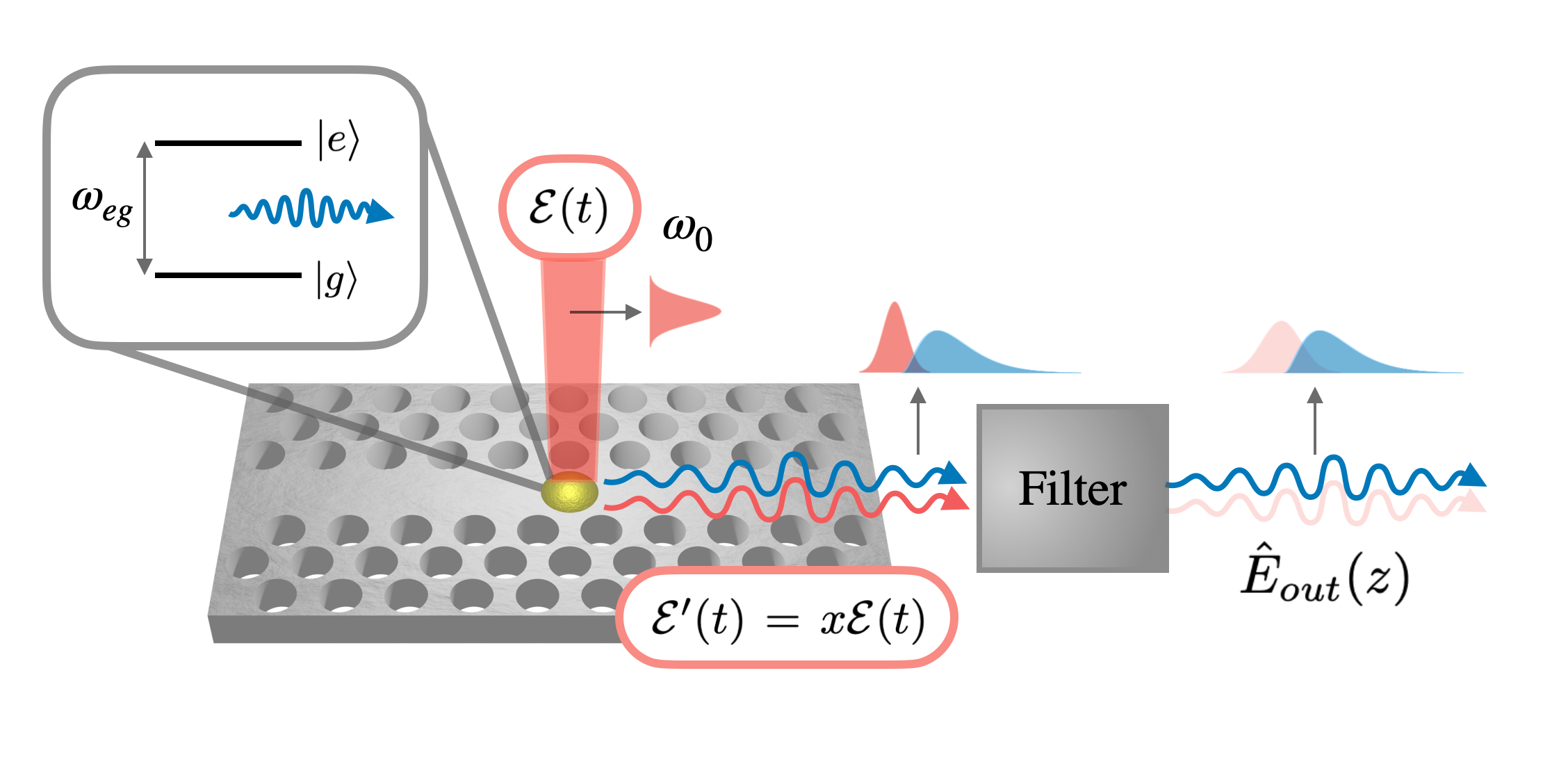}
{ \caption{
A quantum emitter (yellow semisphere) is represented by a two-level system of resonance frequency $\omega_{eg}$ with ground state $\ket{g}$ and excited state $\ket{e}$. The emitter is placed in a waveguide and excited from the top with a coherent field $\mathcal{E}(t)$ of frequency $\omega_0$ that leaks into the waveguide resulting in a field of amplitude $\mathcal{E}'(t)=x\mathcal{E}(t)$, where $x=|x|e^{-i\theta}$ is the leakage factor.  
This field is mixed with the field of the emitter, which consists of a superposition of vacuum, a single-photon state and, with low probability, a two-photon state. The total output field is filtered to reduce the amount of laser leakage so that the emission from the emitter dominates the field after the filtering process in the final output field $\hat{E}_{out}(z)$, as desired. 
}\label{fig:waveguide} }
\end{figure*}

\section{System}
We consider a system consisting of a single emitter, represented as an effective two-level system with ground state $\ket g$ and excited state $\ket e$, as depicted in Fig.~\ref{fig:g2_pulsewidth}(a). The emitter has a natural linewidth $\Gamma$ describing the spontaneous emission rate from the excited state.
We aim to perform a perturbative expansion where we allow at most two photons to be emitted from the quantum dot using a recently developed wavefunction ansatz method~\cite{fischer2017signatures, trivedi_fewphoton_2018, heuck_controlledphase_2020, das2019wave}. 
We model the system using a Hamiltonian given by
\begin{equation}
\hat{H}=\hat{H}_F + \hat{H}_{\text{emitter}} + \hat{H}_I\,, 
\end{equation}
with the electromagnetic field Hamiltonian $\hat{H}_F=\int dk \hbar\omega_k \hat{a}_k^\dagger\hat{a}_k$ and bosonic creation (annihilation) operators $\hat{a}_k^{\dagger}$ ($\hat{a}_k$) with wave numbers $k$ and frequencies $\omega_k$.
The free energy of the emitter is given by $\hat{H}_{\text{emitter}} = \hbar \omega_{eg} \hat{\sigma}_{ee}$, with the emitter operator $\hat{\sigma}_{ee} = \ket{e}\bra{e}$ and frequency $\omega_{eg}$ of the corresponding dipole transition $|g\rangle \leftrightarrow |e\rangle$. 
The interaction between the field and the emitter, placed at position $z_0$, is given by~\cite{chang2007single}
\begin{equation}
 \hat{H}_I= - \hbar \int dk \mathcal{G}_k \hat{\sigma}_{eg} \hat{a}_k e^{ikz_0} + \text{H.c.}\,,
\end{equation}
with the coupling strength $\mathcal{G}_k$ between the emitter and the field and the atomic lowering (raising) operators $\hat{\sigma}_{ge}=\ket g \bra e $ ($\hat{\sigma}_{eg}=\ket e \bra g$).

For simplicity, we study the case in which the emitter is placed in a chiral waveguide, such that we only consider photons travelling in a single direction. For more general interactions, e.g., two-sided waveguides, the system can easily be adapted to the present description by a suitable redefinition of the modes \cite{chang2007single,shen_strongly_2007,witthaut_photon_2010}. In the end, it is only the ratio between the field from the emitter and the leaked field that matters, and the results are therefore independent of this choice of geometry.

We further assume that the relevant frequency components of the field are centered in a narrow frequency interval around the frequency $\omega_0$ with corresponding wavenumber $k_0$, so that we can expand the frequency to lowest order $\omega_k\approx \omega_0+v_g (k-k_0)$, where $v_g=\partial\omega/\partial k $ is the group velocity. 
To describe the light-matter interaction, we also assume that the coupling strength $\mathcal{G}_k$ is similar for all relevant modes $k$ ($\mathcal{G}_k\approx\mathcal{G}$). We transform the Hamiltonian to a rotating frame with respect to the driving frequency $\omega_0$ through $\hat{\tilde{H}}=\hat{U}^{\dagger}\hat{H}\hat{U}$ with the unitary operator
\begin{equation}
  \hat{U} = \exp{-i\left(\hbar\omega_0\hat{\sigma}_{ee} + \hbar\omega_0\int dk \hat{a}^{\dagger}_k\hat{a}_k\right)t} \,.
\end{equation}
We thus obtain the rotating Hamiltonian $\hat{\tilde{H}}$
\begin{multline}
  \hat{\tilde{H}} = \hbar\Delta\sigma_{ee} + \hbar \int dk (\omega_k-\omega_0)\hat{a}_k^{\dagger}\hat{a}_k\\- \hbar \mathcal{G} \int dk \left(\hat{\sigma}_{eg} \hat{a}_k e^{ikz_0} + \text{H.c.}\right)\,,
\end{multline}
where we have defined the detuning with respect to the driving field as $\Delta = \omega_{eg}-\omega_0$.
The interaction between light and matter can conveniently be expressed by the slowly-varying field operator
\begin{equation}
    \hat{E}(z)=\frac{1}{\sqrt{2\pi}}\int dk \hat{a}_{k}e^{ i\left(k-k_0\right)z}\,,   
\end{equation}
which describes the annihilation of photons at position 
$z$. These field operators obey the usual commutation relations $\left[ \hat{E}(z),\hat{E}^\dagger(z') \right]=\delta(z-z')$ and allow us to express the total Hamiltonian in the rotating frame as
\begin{multline}
  \hat{\tilde{H}} = \hbar\Delta\sigma_{ee} -i\hbar v_g\int dz \hat{E}^{\dagger}(z)\frac{\partial{\hat{E}(z)}}{\partial{z}}\\-\hbar\sqrt{2\pi}\mathcal{G}\left(\hat{\sigma}_{eg}\hat{E}(z_0)e^{ik_0z_0}+\text{H.c.}\right)\,.
  \label{eq:hamiltonian_t}
\end{multline}

Ideally, the classical drive which excites the single-photon source is completely separated from the collection of the emitted photons, so that the emission from the source is not contaminated by the drive. This can in principle be achieved by driving a waveguide-coupled emitter from the side such that there are no initial excitations in the waveguide mode that we later use for collection (formally this can be described by an additional coupling channel with a vanishing coupling constant but a large amplitude). With resonant driving, however, there will always be some leakage of the driving field into the detection path, e.g., due to scattering from impurities around the emitter. 
In this case, a fraction of the coherent input field $\mathcal{E}'(t)$ should be included in the output field $E_{\text{out}}$. Here we are interested in understanding the effect of this leaked field on the quality of the single photon source.

To give a proper treatment of the incident field, we consider the initial state of the system to be a tensor product of a coherent photon state and the emitter ground state
\begin{equation}
  \ket {\psi(t=0)}=\hat{D}(\alpha_k)\ket{ \varnothing } \otimes \ket{g} \,,
\end{equation}
with the displacement operator $\hat{D}(\alpha_k)=\exp{\int dk (\hat{a}^\dagger_{k}\alpha_k-\hat{a}_{k}\alpha^*_k)}$ and photon vacuum state $\ket{\varnothing}$.
Instead of applying the displacement to the state, it is convenient to instead transform the rotating frame via the Hamiltonian 
\begin{equation}
\hat{\tilde{H}}'=\hat{D}^\dagger(\tilde{\alpha}_k) \hat{\tilde{H}} \hat{D}(\tilde{\alpha}_k) - i\hbar\hat{D}^\dagger(\tilde{\alpha}_k)\frac{d\hat{D}(\tilde{\alpha}_k)}{dt}\,, 
 \label{eq:int_pic}
\end{equation}
with $\tilde{\alpha}_k=\alpha_k e^{-i(\omega_k-\omega_0)t}$, such that the displaced state $\ket{\psi'(t)}=\hat{D}^\dagger(\tilde{\alpha}_k)\ket{\psi(t)}$ satisfies a transformed Schrödinger equation \cite{mollow_purestate_1975} 
$i\hbar\frac{\partial}{\partial t}\ket{\psi'(t)} = \hat{\tilde{H}}'\ket{\psi'(t)}$. While the free energy and field Hamiltonians ($\hat{\tilde{H}}_{\text{emitter}}$ and $\hat{\tilde{H}}_F$) remain identical under the transformation shown in Eq.\eqref{eq:int_pic}, the rotating interaction Hamiltonian $\hat{\tilde{H}}_I$ yields
\begin{equation}
    \hat{\tilde{H}}'_I=
    -\hbar \sqrt{2\pi}\mathcal{G} \hat{ \sigma}_{eg} \left( \hat{E}(z_0)+ \mathcal{E}'(t) \right)e^{ik_0z_0}+\text{H.c.}\,,
  \label{ham1}
\end{equation}
with the coherent input field 
\begin{equation}
\mathcal{E}(t)=\frac{1}{\sqrt{2\pi}}\int dk\alpha_k e^{i\left[(k-k_0)z_0-(\omega_k-\omega_0)t \right]}\,,
\end{equation}
and where $\mathcal{E}'(t)= x\mathcal{E}(t)$ and we define the leakage parameter $x\equiv|x|e^{-i\theta}$. Here the magnitude $|x|<1$ represents the fraction of the driving field that leaks into the waveguide, and $\theta$ is the phase shift that the field acquires through the scattering process at the position of the emitter ($z_0$). With this particular transformation the initial photonic state becomes simply the vacuum state, and the input field can be replaced by a classical drive with Rabi frequency $\Omega(t)=2\sqrt{2\pi}\mathcal{E}(t)\mathcal{G}$, where we will show below that $\mathcal{G}$ can be related to the decay rate $\Gamma$ through $\Gamma= 2\pi \mathcal{G}^2/v_g$. Then by calculating the evolution of the slowly varying operator $\hat{E}(z)$ through Heisenberg's equation of motion, we obtain
\begin{equation}
  \hat{E}_{out}(z) = \hat{E}_{in}(z)
  + \sqrt{\frac{\Gamma}{v_g}}\left(x\frac{\Omega(t)}{2\Gamma} + i \hat{\sigma}_{ge}\right)\,,
  \label{eq:x_heisenberg}
\end{equation}
where we have performed the transformation discussed above for the incoming field according to $\hat E\rightarrow \hat E +\mathcal{E}'(t)$. When evaluating the output field, it is important to remember that we have performed this transformation, as the outgoing field $\hat E_{\text{out}}$ becomes a combination of the field scattered by the emitter in the transformed picture and the incident field.

The parameterization of the scattering of the driving field through $x$ allows us to fully describe its effect on the output field in a very simple way. The particular process that produces the leakage of the driving field can have multiple origins, e.g. due to scattering happening both outside or inside the waveguide. The particular mechanism is in fact typically unknown for each particular experimental single-photon source and varies with the alignment of the setup. 
We assume that both the leakage fraction $|x|$ and the phase $\theta$ are properties of the particular system itself and are therefore independent of the incident light. In many experimental situations the scattering process will be unstable, e.g., if it takes place outside the waveguide, so that the phase will drift from shot to shot of the experiment. In this case, the results that we derive should be averaged over the values in the different experimental runs. If, on the contrary, the leakage is due to scattering in a nanostructure containing the emitter, this drift may be negligible and there will be a definite phase relation between the leaked field and the emitted field. This phase $\theta$ will influence the performance of the single-photon source, as we analyse in detail below.

\begin{figure}[]
 {\includegraphics[width=0.95\columnwidth]{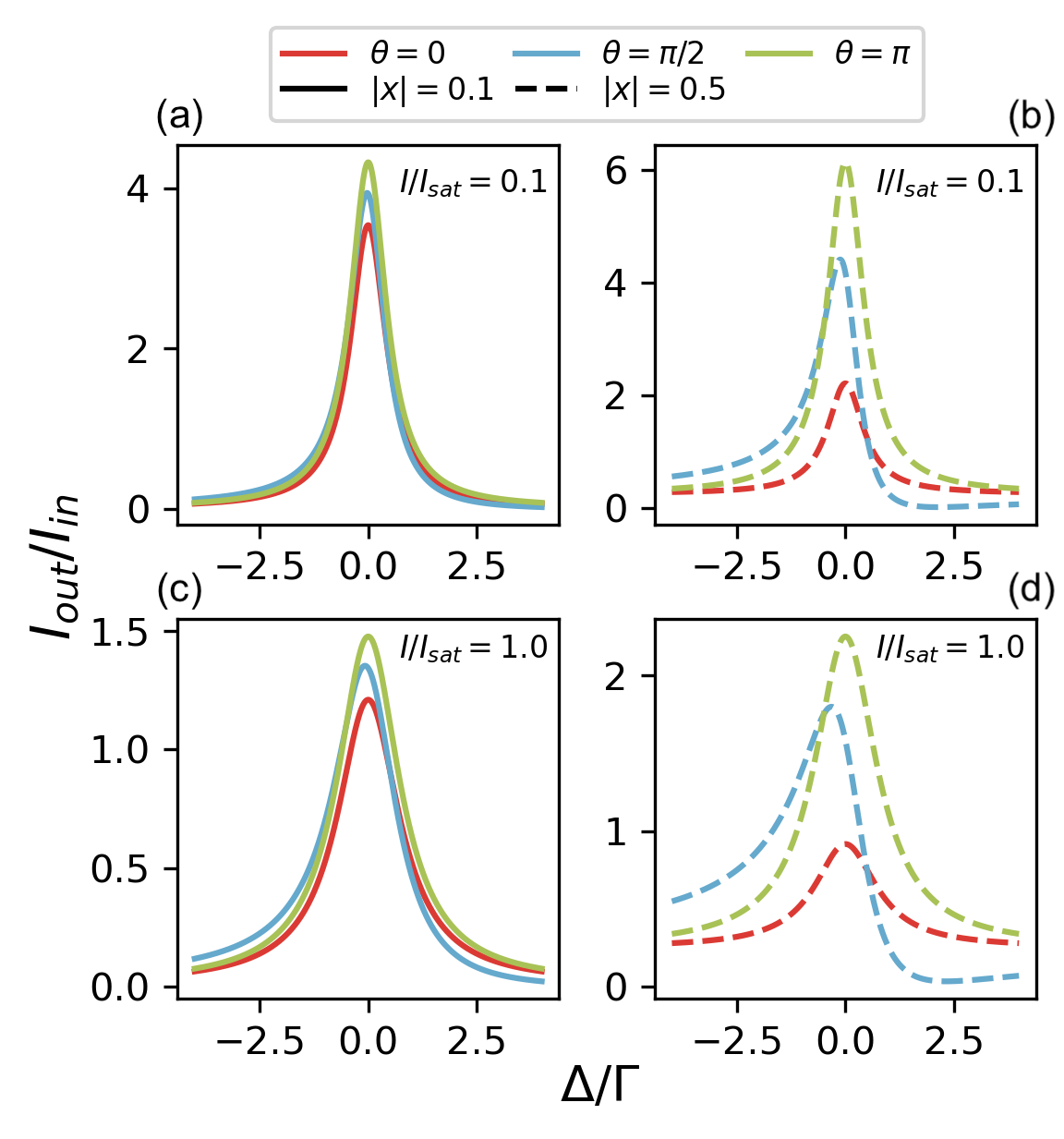}}
{ \caption{Input-output intensity ratio $I_{\text{out}}/I_{\text{in}}$ as a function of the field detuning $\Delta$ for different values of the leakage fraction $|x|$ [$|x|=0.1$ in (a) and (c) and $x=0.5$ in (b) and (d)] and phase $\theta$. We consider CW laser excitation for both weak [ $I/I_{sat}=0.1$ in (a) and (b)], and strong [$I/I_{sat}=1.0$ in (c) and (d)] driving. The incoupling and outcoupling efficiency parameter $A$ has been set to $A=1$ in all cases, which can result in $I_{\text{out}}> I_{\text{in}}$. We observe that the scattering phase induces an asymmetry in the intensity ratio, which allows extracting the leakage fraction $|x|$ and phase $\theta$.
}\label{fig:intensity_detuning} }
\end{figure}

Characterizing the leakage fraction $|x|$ and the scattering phase $\theta$ experimentally through the usual correlation and visibility measurements is not immediately clear from Eq.\eqref{eq:x_heisenberg}. To allow an experimental determination, we calculate the intensity of the output field for a continuous-wave (CW) incident field through the master equation. This allows us
to relate the leakage fraction and phase to experimentally available parameters such as the detuning $\Delta$ of the emitter and the saturation intensity $I_{\text{sat}}$. To this end, we solve the master equation
\begin{equation}
  \dot{\hat\rho}(t) = -i [\hat{\tilde{H}},\hat\rho(t)] + \mathcal{D}(\hat\rho(t))\,,
  \label{eq:master_eq}
\end{equation}
with the Hamiltonian derived in Eq.~\eqref{eq:hamiltonian_t} and
\begin{multline}
  \mathcal{D}(\hat\rho(t)) = \Gamma \hat{\sigma}_{ge}\hat\rho\hat{\sigma}_{eg} + 2\Gamma_d \hat{\sigma}_{ee}\hat\rho\hat{\sigma}_{ee} \\
  - \frac{\Gamma+2\Gamma_d}{2}\left(\hat{\sigma}_{ee}\hat\rho + \hat\rho\hat{\sigma}_{ee} \right)\,. 
\label{eq:liouvillian}
\end{multline}
We have here introduced a pure dephasing rate $\Gamma_d$ in the time evolution of the off diagonal terms of the density matrix since several of the experimental systems under investigation suffer from such dephasing \cite{uppu2020scalable}. From the master equation above, we obtain equations of motion that we can solve to find the steady state solution $\hat\rho = \rho_{ee}\hat{\sigma}_{ee} + \rho_{eg}\hat{\sigma}_{eg} + \rho_{ge}\hat{\sigma}_{ge} + \rho_{gg}\hat{\sigma}_{gg}$ (see Appendix \ref{app:eq_motion_master_eq} for the details). 
With the steady state solution of the master equation and Eq.\eqref{eq:x_heisenberg}, we can finally calculate the measured intensity as the expectation value of the steady state through $I_{\text{out}} = \Tr{v_g\hat{E}_{out}^{\dagger}(z)\hat{E}_{out}(z)\rho}$, obtaining the output intensity
\begin{multline}
I_{\text{out}} = 
\Gamma\rho_{ee} 
  + \frac{1}{2}\left(-ix\Omega(t) \rho_{ge} + \text{H.c.}\right)+ I_{in}\abs{x}^2\,,
\end{multline}
where the input intensity is $I_{in}=\Omega(t)^2/4\Gamma=v_g|\mathcal{E}(t)|^2$. Note that the contribution from the operator $\hat{E}_{in}$ vanishes since the transformed state has the vacuum as the initial state. Inserting the steady state solution from Eq.\eqref{eq:rho_steady}, setting the emitter position at $z_0=0$ and ignoring for simplicity the dephasing $\Gamma_d=0$, the measured intensity finally yields
\begin{widetext}
\begin{equation}
  \frac{I_{\text{out}}}{I_{\text{in}}} = A \left[ \frac{4}{1+\left(\frac{2\Delta}{\Gamma}\right)^2 + 2\frac{I}{I_{\text{sat}}}}\left(1 - (|x|e^{-i\theta}\left(\frac{1}{2}+i\frac{\Delta}{\Gamma}\right)+\text{c.c.})\right)+\abs{x}^2\right] \,,
  \label{eq:intensity}
\end{equation}
\end{widetext}
where the saturation intensity of the system $I_{sat}$ is defined as the intensity where  $\Omega^2=\Gamma^2$ (i.e., $I_{sat}/I=\Gamma^2/\Omega^2$). An efficiency factor $A<1$ is introduced to account for both the incoming laser coupling efficiency and the outcoupling detection efficiency. We choose this factor to be   $A=1$ in Figure \ref{fig:intensity_detuning}, thus artificially allowing $I_{\text{out}}/I_{\text{in}}>1$. Note that this definition does properly reflect the situation when the input laser fully couples to the waveguide ($|x|e^{-i\theta}=1$), since we obtain $I_{\text{out}}=I_{\text{in}}$ as expected. For a complete solution including a dephasing rate $\Gamma_d\neq0$, see Appendix \ref{app:eq_motion_master_eq}. Eq.\eqref{eq:intensity} allows us to obtain both the leakage factor and phase for the system by varying the detuning and incident intensity and fitting the output intensity. 

To obtain an intuitive idea for the typical experimental value expected for $|x|$, we note that for a two level emitter excited with a CW weak laser below saturation (i.e., $I/I_{sat}\ll 1$), we have the ratio
\begin{multline}
 \frac{I_{out}(\Delta \rightarrow \infty)}{I_{out}(\Delta = 0)} =\\
  =1 - \frac{4(1 - \Re{x})}{4(1 - \Re{x}) + |x|^2}\simeq|x|^2/4\,, 
\end{multline}
where we have assumed a low leakage factor $|x|\ll1$. The magnitude of the leakage can thus be found from the ratio of intensities on and off resonance. Due to interference with the field from the emitter (terms proportional to $|x|$), the influence is, however, more complicated than just the addition of a background leakage intensity. We illustrate this in Figure \ref{fig:intensity_detuning}, where we show the intensity as a function of detuning for various values of $|x|$, $\theta$ and $I/I_{\text{sat}}$. 

\section{Equations of motion}
A good single-photon source only emits a single photon at a time and, with very low probability, two. The probability of emitting three photons is therefore expected to be negligible. In our theoretical model we thus employ a weak field scattering description by means of a wavefunction ansatz for the emitter and the fields where we truncate the number of emitted photons to two~\cite{das2019wave, mantasthesis, zhaithesis} (note that we perform the truncation in the frame where the incident field is in the vacuum state; after the inverse transformation of the field, the resulting state may contain more photons). This truncation is valid for a weak excitation field, and/or a short pulse such that the emitter does not have time to emit multiple photons. The latter point will be verified in the following sections, where we calculate the two-photon correlations. The procedure is outlined in Ref.~\cite{das2019wave} but for clarity, we summarize its main features below, where we also clarify various misprints and inconsistencies in the original preprint. 

The wavefunction ansatz contains the joint state of both the fields and the emitter containing up to two excitations
\begin{multline}
  \ket{ \psi'(t)}= c_g(t) \ket{g,\varnothing}+ c_e(t) \ket{e,\varnothing}\\
  + \sqrt{v_g}\int dt_e \phi_{1,g}(t,t_e)\hat{E}^\dagger(v_g(t-t_e)) \ket{g,\varnothing}\\
  + \sqrt{v_g}\int dt_e \phi_{1,e}(t,t_e)\hat{E}^\dagger(v_g(t-t_e)) \ket{e,\varnothing}\\
  + v_g\iint dt_{e2} dt_{e1} \phi_2(t,t_{e2},t_{e1})\\ 
   \cdot \hat{E}^\dagger(v_g(t-t_{e2}))\hat{E}^\dagger(v_g(t-t_{e1})) \ket{g,\varnothing}\,,
   \label{eq:ansatz}
\end{multline}
with $c_g$ and $c_e$ being the amplitudes for the emitter to be in ground and excited states, respectively, with the field being in the vacuum state. The amplitudes $\phi_{1,g}(t,t_e)$ and $\phi_{1,e}(t,t_e)$ correspond to a state with one photon, emitted at a time $t_e$ travelling in the waveguide while the emitter is in the ground state or excited state, respectively. The amplitude
$ \phi_2(t,t_{e2},t_{e1})$ corresponds to two emitted photons, emitted at times $t_{e1}$ and $t_{e2}>t_{e1}$, while the emitter is in the ground state. Since we only consider a maximum of two emissions, we exclude the possibility of the emitter being excited for the two-photon state. 

We then employ the wavefunction ansatz to solve the time dependent Schrödinger equation $i\hbar\frac{\partial}{\partial t}\ket{\psi'(t)} = \hat{\tilde{H}}'\ket{\psi'(t)}$. 
We divide the dynamics into two different time windows: one before the first emission occurs, $0<t<t_{e1}-\epsilon$, and one after $t_{e1}+\epsilon <t <\infty$, where $\epsilon\rightarrow0$. For numerical convenience, the equations of motion are solved numerically in a rotating frame with respect to that used in Eq.~\eqref{eq:eq_motion_app} \cite{das2019wave}. In particular, we introduce the rotated functions $\tilde{c}_e(t) = \exp(i\Delta t)c_e(t)$ and $\tilde{\phi}_{1,e}(t,t_{e,1}) = \exp(i\Delta t)\phi_{1,e}(t,t_{e,1})$. After integrating $\dot{\phi}_{1,g}(t,t_{e,1})$ and $\dot{\phi}_{2}(t,t_{e,2},t_{e,1})$ in the time domain before ($0<t<t_{e,1}-\epsilon$) and after ($t_{e,1}+\epsilon<t<\infty$) the first photon emission respectively, we obtain
\begin{align}
\begin{split}
  \dot{c}_g(t) &= i\frac{\Omega}{2}^*e^{-i\Delta t}\tilde{c}_e(t) \\
  \dot{\tilde{c}}_e(t) &= i\frac{\Omega}{2}e^{i\Delta t}c_g(t)-\frac{\Gamma}{2}\tilde{c}_e(t) \\
  \dot{\phi}_{1,g}(t,t_{e1}) &= i\frac{\Omega}{2}^*e^{-i\Delta t}\tilde{\phi}_{1,e}(t,t_{e1}) \\ 
  \dot{\tilde{\phi}}_{1,e}(t,t_{e1}) &= i\frac{\Omega}{2}e^{i\Delta t }\phi_{1,g}(t,t_{e1}) -\frac{\Gamma}{2}\tilde{\phi}_{1,e}(t,t_{e1}) \,,
  \label{eq:eq_motion}
  \end{split}
\end{align}
along with the boundary conditions
\begin{align}
  \begin{split}
     \phi_{1,g}(t+\epsilon,t) &= i \sqrt{\Gamma} e^{-i\Delta t }\tilde{c}_e(t) , \\ \tilde \phi_{1,e}(t,t) &= 0 ,\\ \phi_{2}(t+\epsilon,t,t_{e1}) &= i\sqrt{\Gamma}e^{-i\Delta t}\tilde{\phi}_{1,e}(t,t_{e1})\,.
     \label{eq:boundary}
  \end{split}
\end{align}
For the initial condition we assume the emitter to be in the ground state $\ket{ \psi'(t=0)}=\ket{g,\varnothing}$ corresponding to $c_g(t=0)=1$. We observe that the equations of motion (Eq.~\eqref{eq:eq_motion}) are simply the usual equations of motion of a driven two-level system with a decaying excited state, as derived e.g. in the quantum Monte Carlo wavefunction approach. These describe the evolution in between decays. The boundary conditions in Eq.~\eqref{eq:boundary} relate the state of the system after a decay, to the state before the decay; e.g., the amplitude to emit a photon at time $t$ and to be in the ground state after the decay $\phi_{1,g}(t+\epsilon,t)$ is proportional to the amplitude of being in the excited state $\tilde{c}_e(t)$ at that time. For further details, see Appendix \eqref{app:eq_motion_master_eq}.

\section{Two-photon correlations for pulsed fields\label{section:two-photon}}
We assume the input field to be a pulsed excitation field. 
In particular, we consider Gaussian excitation $\pi$-pulses 
\begin{equation}
  \Omega(t,\sigma) = \Omega_0 \exp{- \frac{(t-t_0)^2 
  }{2\sigma^2}
  }\,,
  \label{eq: omega_pulse}
  \end{equation}
where $\Omega_0 = \sqrt{\pi/(2\sigma^2)}$, and $\sigma$ is the pulse width.
We choose the normalization $\Omega_0$ so that $\int \Omega(t,\sigma) dt =\pi$ corresponds to a perfect excitation pulse in the short pulse limit, where we can ignore the decay $\sigma\Gamma\ll 1$.
We evaluate the first-order correlation function at time $t$~\cite{das2019wave} 
\begin{multline}
 G^{(1)}\left(t,t_{e1},t_{e2}\right)= v_g \bra{ \psi' (t)} ( \hat{E}^\dagger \left(t_{e1} \right) + \mathcal{E}'^*\left( t_{e1} \right)) \\ \cdot
 ( \hat{E} \left(t_{e2} \right) + \mathcal{E}' (t_{e2}) ) \ket{\psi'(t)} \,,
\label{eq:G1}
\end{multline}
where $t_{e1}$ and $t_{e2}$ the one and two-photon emission times, respectively. 

From the first-order correlation function we can trivially obtain the mean number of photons $\bar n$ over the total duration of the pulse stretching from $t_0$ to $t_T$ 
\begin{equation}
\begin{split}
 \bar n = \expval{n} = \int^{t_T}_{t_0} dt G^{(1)} (t_T,t,t)\,.
\end{split}
\end{equation}
We similarly find the two-photon correlation function
\begin{multline}
 G^{(2)}(t,t_{e2},t_{e1})= v^2_g\sum_{n} \bigg \lvert \bra{n}
 \left( \hat{E} \left(t_{e1} \right) 
 + \mathcal{E}'\left( t_{e1} \right) \right) \\ 
  \cdot\left(\hat{E} \left(t_{e2} \right) + \mathcal{E}' (t_{e2}) \right) \ket{\psi' (t)} \bigg \rvert^2 \,.
  \label{eq:G2}
\end{multline}
Here we have inserted a complete set of states $\ket{n}$ and written the resulting expression as the sum of absolute squares. Exploiting this structure allows us to considerably reduce the numerical complexity of evaluating the expressions, as we explain below. 
In all of these expressions, $t>t_{e1},t_{e2}$ describes some time after the emission of all photons. Since any subsequent evolution of the emitter cannot change already emitted photons, the exact time $t$ is not important for the result. The expression may, however, have different contributions depending on the evaluation time $t$ since e.g. the one photon component oscillates between the emitter being in the ground $\tilde{\phi}_{1,g}$ and excited $\tilde{\phi}_{1,e}$ states. In our numerical simulations, we fix the evaluation time for each choice of parameters, ensuring that the emitter has completely decayed. In addition, the time grid in which the equations of motion from Eq.~\eqref{eq:eq_motion} are solved is also adapted as a function of the evaluation time and the time at which the pulse starts, $t_0$.

\begin{figure}[]
{\includegraphics[width=0.95\columnwidth]{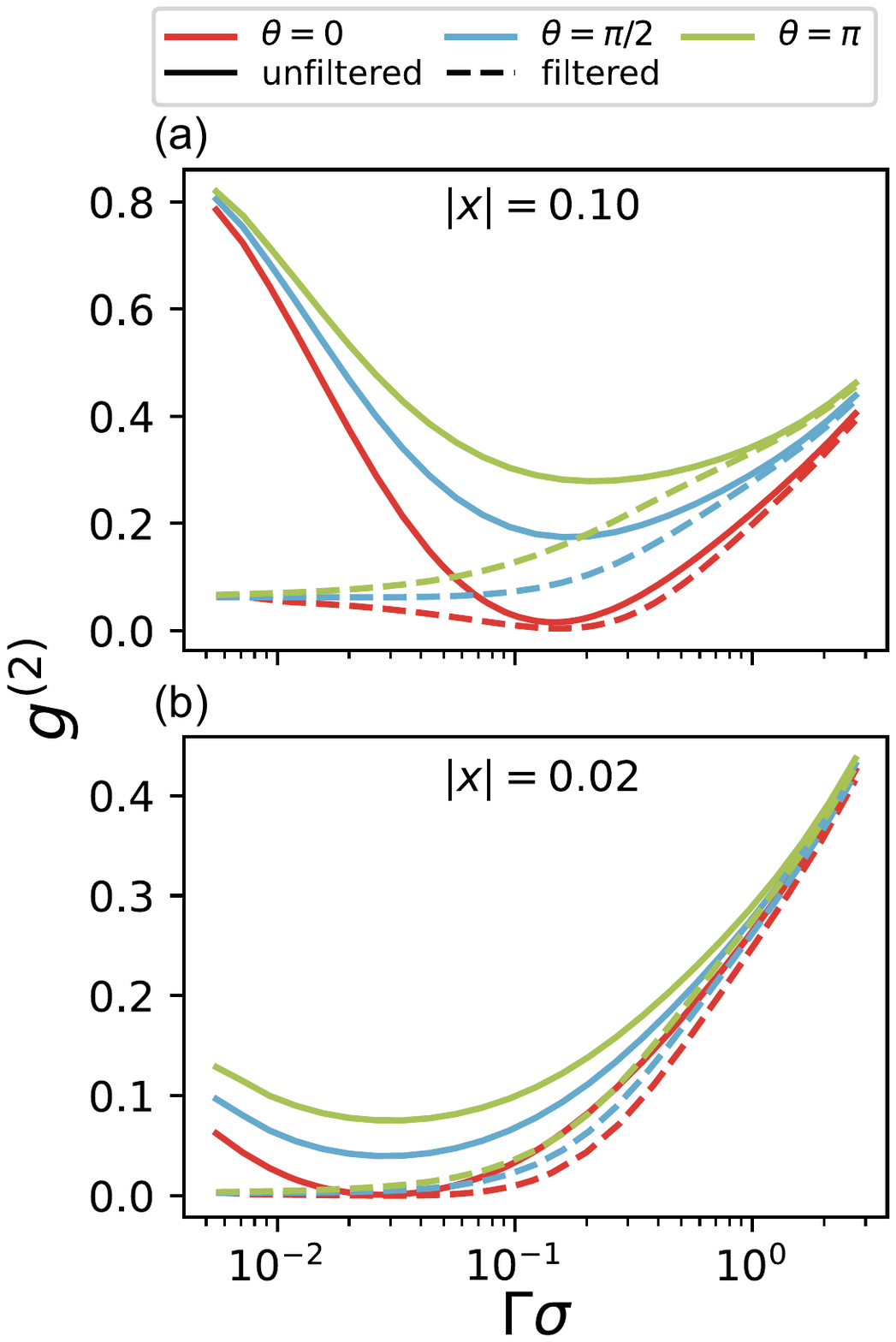}}
\vspace*{-4mm}
{ \caption{
Second-order two-photon correlation, $g^{(2)}(\sigma)$, for various pulse durations $\Gamma\sigma$ of the excitation pulse [the leaked field fraction is $|x|=0.1$ in (a) and $x=0.02$ in (b)]. For short pulses, the leaked excitation field $\mathcal{E}'=|x| e^{-i\theta} \Omega(t)/\sqrt{4\Gamma v_g}$ contains many photons, which has a strong impact on the single-photon purity. 
The leaked field interferes with the emitted field and gives constructive (destructive) interference of the two-photon component for $\theta=\pi$ ($\theta=0$) resulting in higher (lower) values of $g^{(2)}(\sigma)$. For $\theta=0$, interference effects may actually produce lower $g^{(2)}(\sigma)$ for a high $|x|$ as compared to smaller values in some pulse width interval.
Post-emission filtering with a Lorentzian frequency filter of width $\gamma = 1.66\Gamma$ improves the single-photon purity for short pulses, since the leaked field typically has a broad spectral profile in the frequency domain, while the single photon field from the emitter is narrower in frequency. The corresponding filter efficiency in terms of output power is shown in Fig.~\ref{fig:g2_leak_with_inset}.
Large two-photon contributions $g^{(2)}(\sigma) \gtrsim 0.5$ in the long-pulse limit correspond to a regime where the underlying two-photon approximation for the wavefunction ansatz begins to fail due to the increasing contribution from three-photon states. 
}\label{fig:g2_pulsewidth} }
\end{figure}

Expressions for Eqs.~\eqref{eq:G1} and \eqref{eq:G2} in terms of the wavefunction components are given in Appendix~\eqref{app:G1_G2}.
From the correlation functions $G^{(1)}(t,t_{e1},t_{e2})$ and $G^{(2)}(t,t_{e1},t_{e2})$, we calculate the normalized second order correlation function $g^{(2)}$ integrated over the whole excitation pulse duration~\cite{zhaithesis,kiraz2004quantum}
\begin{equation}
  g^{(2)}(t)=\frac{\iint dt_{e1}dt_{e2} G^{(2)}(t,t_{e1},t_{e2})}{ \left(\int dt_{e1}G^{(1)}(t,t_{e1},t_{e1}) \right)^2 }\,.
  \label{eq:g2}
\end{equation}
Fig.~\ref{fig:g2_pulsewidth} shows the simulated photon impurity $g^{(2)}(\sigma)$ of the output light for various leakages and pulse durations. Note that in all the analysis, the time correlations are always evaluated at a time $t$ much longer than the pulse duration and the decay rate of the emitter. Thus in the following, the relevant parameter dependence of the pulsed second order correlation function we denote in the paper is the incident pulse width $\sigma$, omitting the time dependence $t$. In particular, note that the dependence on the pulse duration $\sigma$ is clearly non-monotonic for all phases $\theta$ and leakages $|x|$.

It should be noted that for large pulse durations, on the order of the lifetime of the emitter $\sigma\sim 1/\Gamma$, the multi-photon emission $g^{(2)}(\sigma)$ increases due to the higher probability of multiple excitations of the emitter during the duration of the pulse. Therefore, our wavefunction ansatz begins to break down, since we truncated it at a maximum of two  emissions. In the short-pulse limit, on the other hand, the impurity $g^{(2)}(\sigma)$ decreases until it rapidly increases again. 
This effect has been observed in experiments~\cite{ollivier2021hong}, and is caused by the increased pulse power required to produce a short $\pi$-pulse~\cite{giesz2016coherent}. The relative power of the leaked field scales with $|x|^2/\sigma\Gamma$ in the total output so that for short pulses the leaked field completely dominates the total photon count. Since the leaked field is a coherent state with $g^{(2)}(\sigma)=1$, the second order correlation functions will grow towards $g^{(2)}(\sigma)=1$ for sufficiently short pulses (which we have verified in our simulations). 

For $|x|\lesssim0.1$ the short-pulse and long-pulse limits discussed above can be viewed as limiting cases where the leaked field dominates or can be neglected, respectively.
In the intermediate regime, however, interferences between the leaked field and the field emitted from the emitter also play an important role. This is evident from the comparatively large impact of the phase parameter $\theta$. Note that  we consider the excitation to the excited state $\ket{e}$ from which the light is emitted into a Fock state with no well-defined phase.  The emission of two photons from the emitter, however, requires that one of them is emitted during the pulse such that the emitter can be re-excited. During the pulse,  the emitter is in a superposition of ground and excited states and the emission thus has a well-defined phase and can interfere with the leaked field.  Since the precise cause of the leaked field varies from experiment to experiment, we explore different values of the phase in order to understand its influence on the results. We consider three values, $\theta=0$, $\theta=\pi/2$, and $\theta=\pi$ to explore the full range of possibilities (for resonant driving, all observables are symmetric with respect to $\pm \theta$). 

Since the field of the emitter acquires a phase of $-1$ from the absorption and re-emission, c.f. Eqs. \eqref{eq:eq_motion} and \eqref{eq:boundary}, a phase $\theta\approx 0$ corresponds to destructive interference between the two fields. Due to the destructive interference, both $g^{(2)}(\sigma)$ and HOM visibilities (discussed below) can behave in somewhat unexpected manners. In general, $\theta=0$ leads to the lowest two-photon contribution of the three considered phases due to the destructive interference. For some pulse durations, we find that an increase of the leakage fraction $|x|$ can even decrease the two-photon component of the output field (not directly observable in Fig.~\ref{fig:g2_pulsewidth}, which only considers fixed $|x|$). This is unexpected since an increase in $|x|$ means a higher contribution from the leaked field, which has a larger two-photon component than the emitter field. 
This is a curious interference effect, which would be very interesting to look for in experiments. 

On the contrary, $\theta=\pi$ corresponds to constructive interference between the field emitted from the emitter and the leaked field. This consequently corresponds to a much larger two-photon contribution $g^{(2)}$ for all pulse durations and leakage fractions.

For the intermediate angle $\theta=\pi/2$, the leaked field is purely imaginary. When we calculate the absolute value of the field to find the intensities, there is thus limited interference with the purely real field for resonant driving. As a consequence, the result represents an approximate average of all possible angles and we expect it to roughly match the situation where the phase is drifting from shot to shot of the experiment, although these situations are not exactly identical \cite{wenniger_quantum_2024}. For quantities like $g^{(2)}(\sigma)$ and HOM visibility,
the calculated values corresponding to $\theta=\pi/2$ typically fall between the values of $\theta=\pi$ and $\theta=0$. 
\begin{figure}[] \includegraphics[width=\columnwidth]{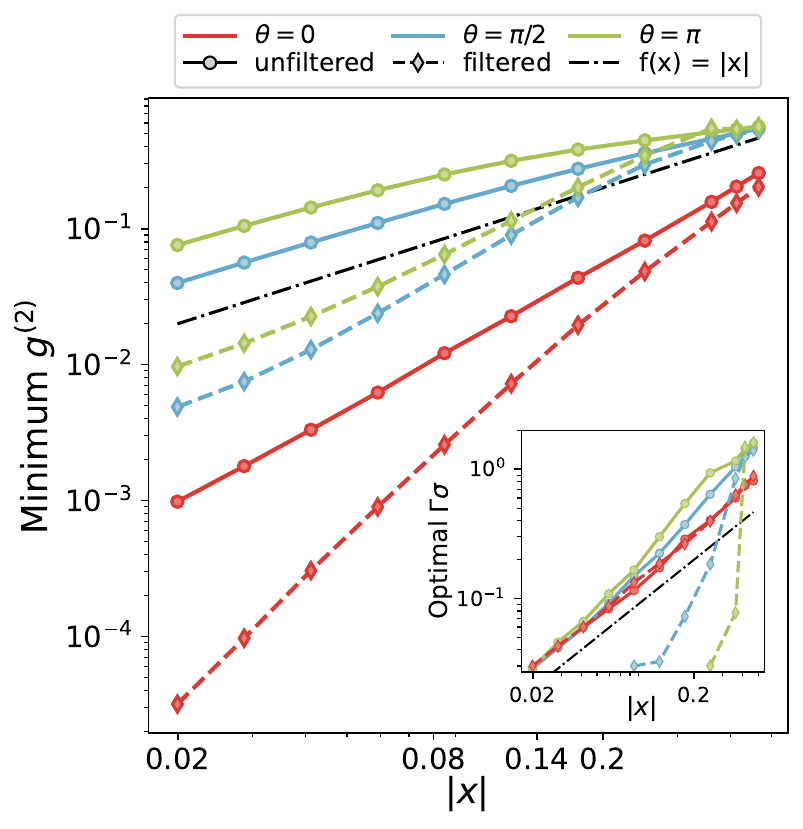}
\caption{
Minimum two-photon contribution of the single-photon source as a function of the leakage amplitude $|x|$. 
The lowest attainable two-photon emission probability of the source increases with the size of the leakage $|x|$. The best results are obtained for leaked fields which have destructive interference with the field emitted from the emitter $\theta=0$. Full curves corresponds to the unfiltered case, while dashed curves corresponds to a Lorentzian frequency filter of width $\gamma=1.66\Gamma$. The dash-dotted line shows $f(x)=|x|$ to indicate the scaling. 
Inset: Corresponding optimal pulse duration $\sigma$ for minimum two-photon emission, as a function of leakage amplitude $|x|$. With lower leakage fraction $|x|$, we have less influence of the leaked field, and it is thus desirable to work with a shorter pulses duration to eliminate multiple excitations of the emitter. For some of the results for the filtered case, the minimum is at $\sigma\rightarrow 0$. 
We note that due to finite numerical accuracy, for very small values of $g^{(2)}$, there may be a small absolute but large relative uncertainty associated with these points.}\label{fig:g2_leak_with_inset} 
\end{figure}
%%%%%%%%%%%%%%%

We now proceed to discuss how the minimum value of $g^{(2)}(\sigma)$ depends on the leakage fraction at various phase values. As shown in Fig.~\ref{fig:g2_leak_with_inset}, the pulse length for which the minimum occurs increases monotonously with the leakage fraction $|x|$ for all phases. This behaviour can be understood by noting that the minimum value of $g^{(2)}(\sigma)$ is obtained at the crossing point between being dominated by the two-photon emission due to the re-excitation of the quantum dot and the contribution due to laser leakage. The former happens with a probability proportional to $\sigma\Gamma$, while the latter is proportional to $|x|^2/\sigma\Gamma$. Thus, one expects the minimum value to occur at a time proportional to $|x|/\Gamma$, which is consistent with the behaviour in the inset of Fig.~\ref{fig:g2_leak_with_inset}. By the same argument, the minimum value of $g^{(2)}(\sigma)$ is also expected to be proportional to $|x|$ in the absence of filtering. As seen in the figure, this is approximately the case except for $\theta=0$. In this case, the destructive interference between the two fields yields a much lower value of the two-photon component. 

\section{Post-emission filtering}
We now consider the effect of post-emission spectral filtering of the output field. This is done via a linear map for the field operators in the Heisenberg picture~\cite{gardiner1985input} 
\begin{equation}
\begin{split}
  \hat{E}_{\text{F}}\left( \omega \right)&= \mathcal{T} \left( \omega - \omega_c \right)\hat{E}\left( \omega \right)\\ &\rightarrow\mathcal{T} \left( \omega - \omega_c \right) (\hat{E}\left( \omega \right)+ \mathcal{E}'\left( \omega \right) )\,,
  \label{eq:Heis_filter}
\end{split}
\end{equation}
where $\mathcal{T}(\omega)$ is the transmission function of the filter, and $\omega_c$ is the central frequency. Here, the first (second) line is valid before (after) the transformations in Eq.~\eqref{eq:int_pic}. For brevity, we omit the vacuum noise operators, which must be present if the transmission is below unity. Since all the quantities we consider are written as normal-ordered products, such vacuum operators will have a vanishing contribution.
We go back to the time domain using
\begin{equation}
  \hat{E}_{\text{F}}\left( t \right)=\int d\omega e^{-i\omega t}\hat{E}_{F}\left( \omega \right)\,,
\end{equation}
and proceed to evaluate the correlation functions (Eqs. \eqref{eq:G1} and \eqref{eq:G2}) via the filtered field mode operators in the time domain (for more details, see Appendix \ref{app:filtering}).

Our analysis can be employed for any kind of filter, which is a major advantage of the wavefunction ansatz method identified in Ref.~\cite{das2019wave}. In this work, we choose a Lorentzian filter 
\begin{equation}
    \mathcal{T}\left( \omega-\omega_c \right)=\frac{\gamma}{i\left( \omega-\omega_c \right) - \gamma}\,,
\end{equation}
where $\gamma$ relates to the full width at half maximum (FWHM) of the filtering function through FWHM = $2\gamma$. 
This choice is motivated by the fact that it describes the effect of a filter cavity. For this particular filter, 
results can also be obtained by including a filter cavity and simulating the density matrix of the joint system of emitter and cavity using the quantum regression theorem. We have explicitly verified that this reproduces all the results for some parameters, but since both methods are computationally demanding, we have not reproduced them for the entire parameter space.

\begin{figure}[ht]
{\includegraphics[width=0.95\columnwidth]{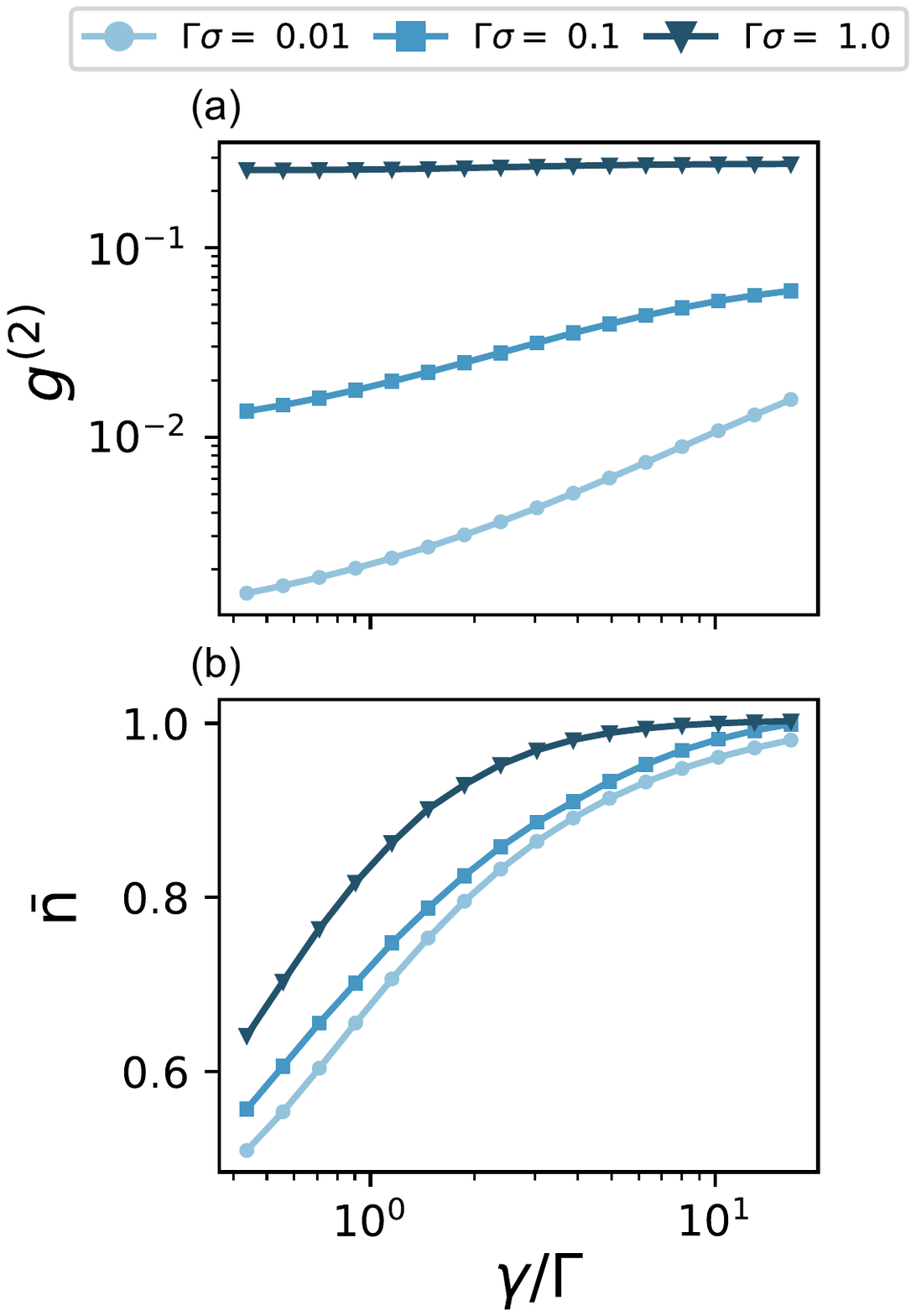}}
\vspace*{-4mm}
 \caption{Two-photon contribution and filtered output photon number for different filter widths $\gamma/\Gamma$ and $\theta=\pi/2$. All results are plotted for $|x|=0.02$. (a) Two-photon contribution $g^{(2)}$ at various pulse durations. A narrow filter
 clearly suppresses the two-photon contribution, in particular in the short-pulse limit. 
 (b) Filtered output photon number $\bar n$. We note that the narrow filter reduces the output as expected. The filtered photon number depends only weakly on pulse length in the short-pulse limit, indicating that much of the leaked field, which is broad in frequency, is filtered out. On the other hand, the single photon contribution is independent of pulse durations and thus only weakly affected.
 }\label{VHOM_pulsewidtha} 
\end{figure}

In Figures ~\ref{fig:g2_pulsewidth} and~\ref{fig:g2_leak_with_inset} we also present results after the application of the frequency filter. 
The effects of the low-pass filter are evident: it reduces high-frequency components, which are mainly present for very short pulses. As a consequence, the frequency filter is particularly efficient in improving the multi-photon component in the short pulse regime. 
As noted in the previous section, the output power of the leaked field scales with $|x|^2/\sigma\Gamma$, so for short pulses this field dominates and is increasingly diminished by the filter. 
Furthermore, a low-pass filter has little effect in the long-pulse limit.

The filter effect on $g^{(2)}(\sigma)$ is striking, with the two-photon component being significantly reduced in the short pulse limit. This is simply a consequence of the leaked field being filtered post emission, while the emitter output is largely unfiltered.
This effect is emphasized for smaller leakage fractions $|x|$, since the turning point of $g^{(2)}(\sigma)$ is pushed towards shorter pulses. This means a greater separation between the frequency profile of the emitter-field and the leaked field, so that the leaked field is more easily filtered out. In Fig.~\ref{fig:g2_pulsewidth} we can see that for leakage $|x|=0.02$ and phases $\theta\gtrsim \pi/2$ the filter completely diminishes the minimum of $g^{(2)}(\sigma)$, so that the two-photon component is monotonously decreasing with short-pulses and reaches a minimum for $\sigma\rightarrow 0$. 

A drawback of applying the filter, however, is that it also diminishes the single photon component that we are interested in. We explore this in Fig.~\ref{VHOM_pulsewidtha} where we plot $g^{(2)}$ and $\langle n\rangle $ as a function of filter width for different pulse durations. As shown in the figure, the filter has only limited effect on the photon number for broad filters $\gamma \gg \Gamma$, but can still give a significant reduction in $g^{(2)}$, especially for short pulses.

\section{HOM Visibility}
The HOM visibility $V$ of a single-photon source provides information about the degree of indistinguishability of the single photons. In addition to the indistinguishability, the HOM visibility is, however, also affected by the multi-photon emission. Extracting the indistinguishability thus requires knowing the effect of the multi-photon emission, which is non-trivial as shown in Refs. \cite{ollivier2021hong,sekatski2022}. In the present theoretical study, the state of the emitter light is pure, meaning that the intrinsic indistinguishability of the single-photon component is set to unity in the shown results. Any reduction in HOM visibility that we find, is thus only due to the two-photon component and this allows us to investigate the relation between the impurity and the HOM visibility. Nevertheless, imperfect intrinsic indistinguishability is accounted for in our analysis as shown below, further detailed in Appendix \ref{app:flimits_reducedV}.

We start by using our wavefunction ansatz to calculate the HOM visibility as a function of the coherence functions introduced previously. This allows us to relate the visibility and the multi-photon emission of the source by means of $g^{(2)}$ and study how it is influenced by the excitation pulse length.
In order to calculate the HOM visibility, we consider a 50:50 beam splitter with input modes $\hat{a}_1$ and $\hat{a}_2$, and output modes $\hat{a}_3$ and $\hat{a}_4$. 
The probability of detecting coincidence counts $P_{cc}$ in both output ports simultaneously is
\begin{equation}
  P_{cc} = \int dt_edt'_e\expval{\hat{a}^{\dagger}_3(t_e)\hat{a}_3(t_e)\hat{a}^{\dagger}_4(t_e')\hat{a}_4(t_e')}\,.
  \label{eq:P_cc}
\end{equation}
For simplicity we only consider the typical experimental scenario: the low efficiency limit where most of the photons are lost before the detectors. In this case, we can simply evaluate the coincidence probability by the two point correlation function. For sources with high efficiency, however, one would need to consider what happens if three or more photons arrive at the detector simultaneously \cite{sekatski2022}. In the low efficiency limit, such events are negligible so that we can restrict ourselves to the expression in Eq.~\eqref{eq:P_cc} and use it to evaluate the visibility 
\begin{align}
\begin{split}
  V = 1 - \frac{P_{cc}}{P_{cc,\text{dist}}}\,,
\end{split}\label{eq:HOMvis}
\end{align} 
where $P_{cc,\text{dist}}$ is the maximum coincidence counts probability achievable, that is, for completely distinguishable photons.
Applying the beam splitter transformations
\begin{equation}
\hat{a}^{\dagger}_1 \rightarrow \frac{1}{\sqrt{2}}\left( i \hat{a}^{\dagger}_3 + \hat{a}^{\dagger}_4\right), \quad
\hat{a}^{\dagger}_2 \rightarrow \frac{1}{\sqrt{2}}\left( \hat{a}^{\dagger}_3 + i\hat{a}^{\dagger}_4\right)\,,
\label{eq:bs_transf}
\end{equation}
and the usual commutation relations $[\hat{a}_i(t_e),\hat{a}^{\dagger}_j(t_e')]=\delta_{ij}\delta(t_e-t_e')$ to Eq.\eqref{eq:P_cc}, we obtain the probability of coincidence counts as a function of the input operators
    \begin{equation}
    \begin{split}
    P_{cc} = \frac{1}{4}\int dt_edt'_e(&\expval{\hat{a}^{\dagger}_1(t_e)\hat{a}^{\dagger}_1(t_e')\hat{a}_1(t_e')\hat{a}_1(t_e)}\\
    +&\expval{\hat{a}^{\dagger}_2(t_e)\hat{a}^{\dagger}_2(t_e')\hat{a}_2(t_e')\hat{a}_2(t_e)}\\
    +2&\expval{\hat{a}^{\dagger}_1(t_e)\hat{a}_1(t_e)\hat{a}^{\dagger}_2(t_e')\hat{a}_2(t_e')}\\
    -2&\expval{\hat{a}^{\dagger}_1(t_e)\hat{a}^{\dagger}_2(t_e')\hat{a}_1(t_e')\hat{a}_2(t_e)})\,.
    \end{split}\label{eq:pcc_1}
    \end{equation}
The first two terms are related to the second order correlation function $G^{(2)}$ of the input fields, while the last two are connected to the first order correlation function. We thus have
\begin{align}
  \begin{split}
    P_{cc} &= \frac{1}{4}\iint dt_e dt_e'(G_1^{(2)}(t_e,t_e')+G_2^{(2)}(t_e,t_e')\\
    &+2G_1^{(1)}(t_e,t_e)G_2^{(1)}(t_e',t_e')-2G_1^{(1)}(t_e,t_e')G_2^{(1)}(t_e',t_e))\,.
  \end{split}
  \label{eq:pcc}
\end{align}
Note that the last term in Eq.~\eqref{eq:pcc} contains the level of indistinguishability of the photons, while the others refer to the usual one and two-photon number correlations.

Following Eq.~\eqref{eq:HOMvis}, we normalise $P_{cc}$ through the probability $P_{cc,\text{dist}}$ of detecting coincidence counts when the photons are completely distinguishable, i.e., by taking Eq.~\eqref{eq:pcc} in the limit where we omit the last term
\begin{align}
  \begin{split}
    P_{cc,\text{dist}} &= \frac{1}{4}\iint dt_e dt_e'(G_1^{(2)}(t_e,t_e')+G_2^{(2)}(t_e,t_e')\\
    &+2G_1^{(1)}(t_e,t_e)G_2^{(1)}(t_e',t_e'))\,.
  \end{split}
  \label{eq:pmax}
\end{align}

We assume the single-photon source to be the same for both the input ports of the beam splitter (meaning $G_1^{(1)}=G_2^{(1)}\equiv G^{(1)}$ and $G_1^{(2)}=G_2^{(2)}\equiv G^{(2)}$). From Eqs. \eqref{eq:pcc} and \eqref{eq:pmax} we thus obtain the HOM visibility
\begin{equation}
  \begin{split}
    V = \frac{\iint dt_e dt_e' \abs{G^{(1)}(t_e,t_e')}^2}{\iint dt_e dt_e' \left( G^{(2)}(t_e,t_e')+G^{(1)}(t_e,t_e)G^{(1)}(t_e',t_e')\right)}\,.
  \end{split}
  \label{eq:final_VHOM}
\end{equation}
From this expression, we can now evaluate the HOM visibility using the previously found correlation functions.

\begin{figure}[ht]
 {\includegraphics[width=0.95\columnwidth]{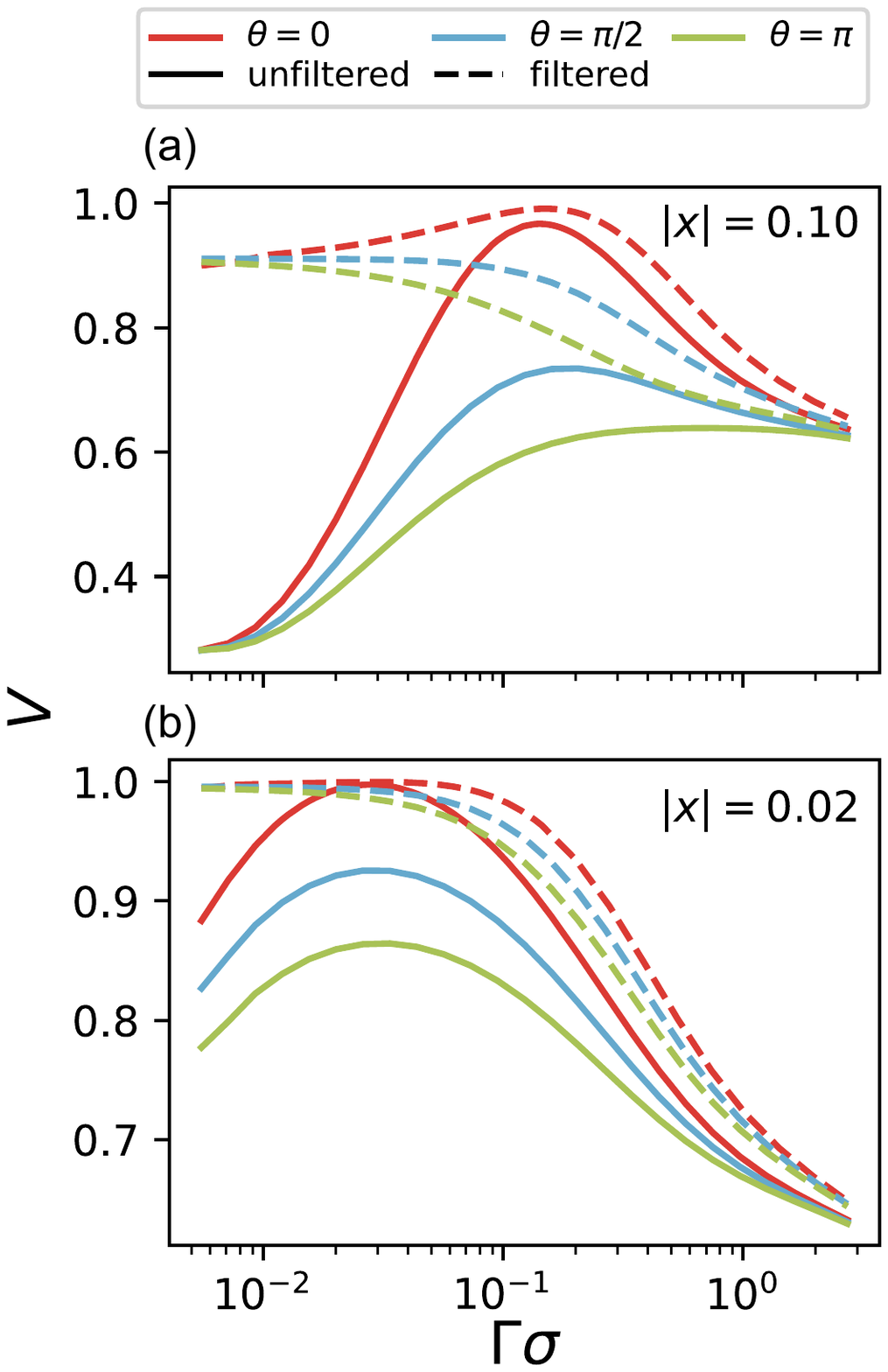}}
 \vspace*{-5mm}
{ \caption{HOM visibility of a single photon source subject to leakage of the driving field with amplitude $|x|$ and phase $\theta$ [$|x|=0.1$ in (a) and $x=0.02$ in (b)]. The filtered values correspond to applying a Lorentzian frequency filter of width $\gamma = 1.66\Gamma$. The visibility peaks at intermediate pulse lengths where $g^{(2)}$ has a minimum, c.f. Fig.~\ref{fig:g2_pulsewidth}.
}\label{VHOM_pulsewidth} }
\end{figure}

 In Fig.~\ref{VHOM_pulsewidth} the HOM visibility $V$ as a function of pulse length is plotted for different leakage factors $|x|$ and phases $\theta$. As seen in the figure, $V$ reaches a maximum at pulse lengths near the minima in $g^{(2)}(\sigma)$. This is due to the two-photon contributions to the output field diminishing $V$, both in the long-pulse and short-pulse regime. 
The leakage has an adverse effect on the visibility in the short-pulse limit, where it decreases the visibility due to pollution from the leaked field (as shown in Fig.~\ref{fig:g2_pulsewidth})). For $\theta=\pi/2$ and $\theta=\pi$, a larger leakage gives a lower visibility at all pulse durations. Like for $g^{(2)}$, however, in the special case $\theta=0$ interference may actually cause the leakage to enhance the visibility within a given region of pulse lengths (not directly visible in the figure). 
At all pulse lengths and values of $|x|$ the filtering enhances visibility, but most noticeably in the short-pulse limit. Furthermore, the best (worst) visibility is always obtained for $\theta=0$ ($\theta=\pi$), where there is destructive (constructive) interference of the leaked field and two photon emission from the emitter. 

\begin{figure}[]
 {\includegraphics[width=0.9\columnwidth]{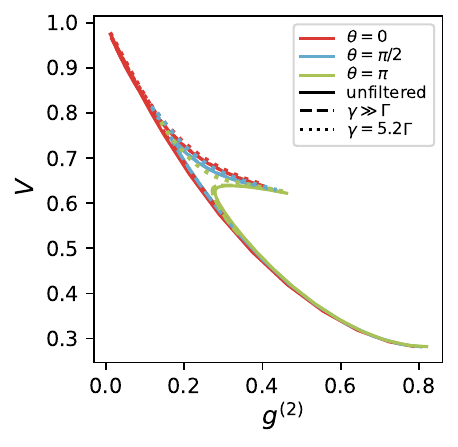}}
 \vspace*{-3mm}
{ \caption{Relation between HOM visibility $V$ and $g^{(2)}$ with leakage $|x|=0.1$. In the curves, we vary the scattering phase $\theta$ and the filtering width $\gamma$, and plot the corresponding values of $g^{(2)}$ and $V$. 
The functions are multivalued, since the visibility is reduced both in the long- and the short-pulse limit. The turning point corresponds to the maximum visibility and the minimum in $g^{(2)}$. 
}\label{fig:Vhom_g2} }
\end{figure}

\begin{figure}[]
 {\includegraphics[width=\columnwidth]{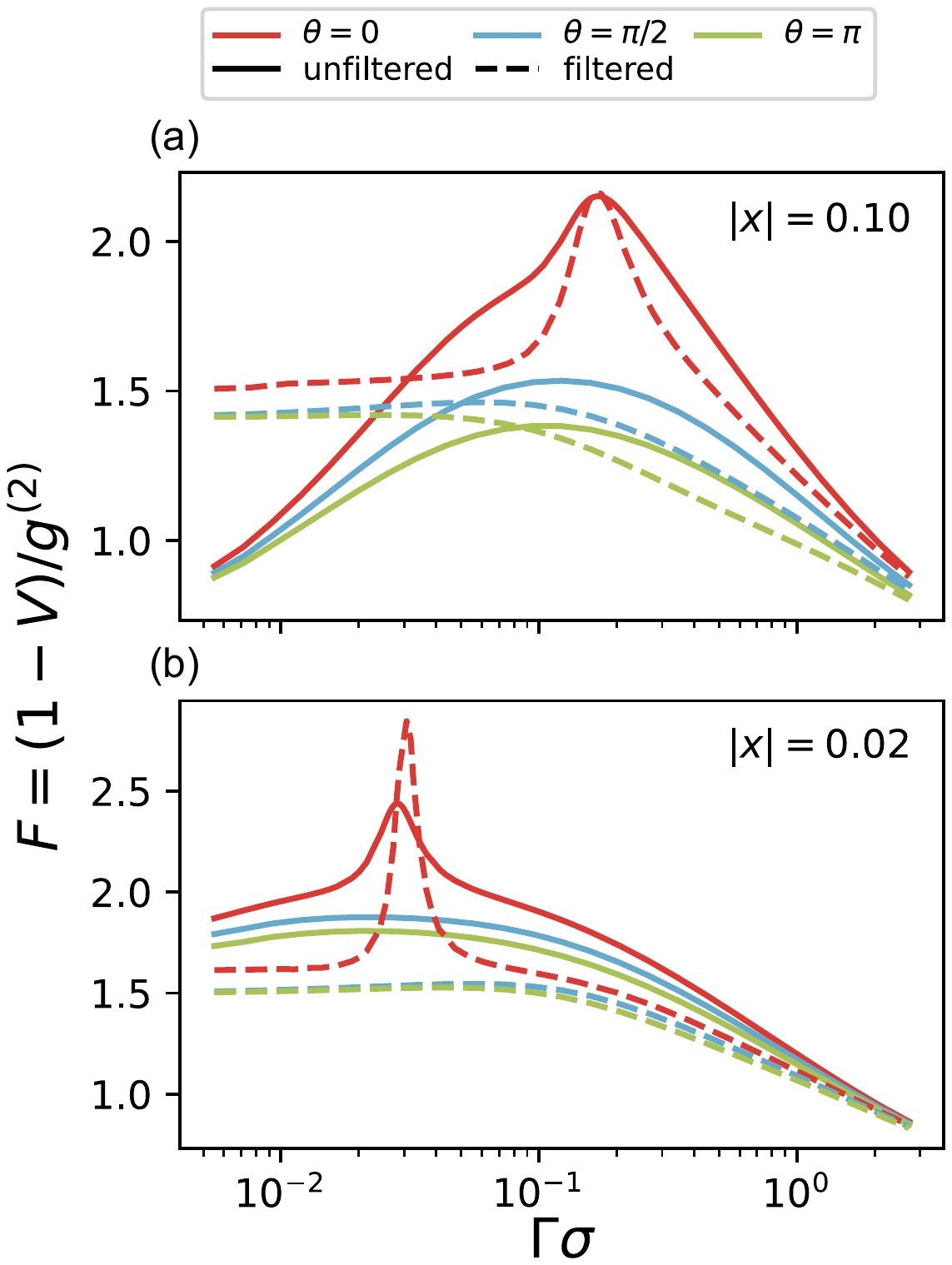}}
 \vspace*{-10mm}
{ \caption{Ratio between lack of visibility and multi-photon emission $F=(1-V)/g^{(2)}$ at different pulse lengths. The filtered values correspond to applying a Lorentzian frequency filter of width $\gamma = 1.66\Gamma$ and the leaked field fraction is set to $|x|=0.1$ in (a) and $x=0.02$ in (b). A high value of $F$ indicates that the two-photon component differs from the mode of the single photon component. The highest value is therefore achieved for $\theta=0$, where the leaked field and the emitter emission interfere destructively, which tends to eliminate the part of the two-photon component that resembles the single photon state. 
}\label{fig:F_factor} }
\end{figure}

In Fig.~\ref{fig:Vhom_g2} we explore the relation between the visibility $V$ and multi-photon emission $g^{(2)}$ for various pulse durations. As seen in the figure, a larger value of $g^{(2)}$ tends to decrease the visibility $V$. It is seen, however, that $V$ decreases more rapidly with $g^{(2)}$ towards the short-pulse limit than the long pulse limit. This indicates that the two-photon components from the leaked field are more detrimental to $V$ than the two-photon components in the long-pulse limit, which originate from the re-excitation of the emitter. This can be understood by noting that in the short-pulse regime, the second photon emitted comes directly from the leaked excitation laser, and is thus (partially) distinguishable from the single photon coming from the emitter. As we discuss in detail below, distinguishable photons diminish $V$ more than indistinguishable photons.
At even shorter pulses, the leaked field completely dominates, so that both single-photon and two-photon components largely originate from the leaked field $\mathcal{E}$. In this limit all photons become indistinguishable and we expect to reach the values $g^{(2)}=1$ and $V=1/2$ of a coherent state.
In the figure we furthermore observe that the relation between $V$ and $g^{(2)}$ is only weakly affected by the value of $|x|$ and filtering, but lower values of $|x|$ allow reaching a more ideal single photon source with lower $g^{(2)}$ and higher visibility. 

To further investigate the relation between the two-photon correlations and visibility, we derive a relation between them in Appendix \ref{app:flimits_reducedV}. In this derivation we decompose the density matrix $\hat\rho=P_0\hat\rho_0+P_1\hat\rho_1+P_{n>1}\hat\rho_{n>1}$ into components $\hat\rho_0$, $\hat\rho_1$ and $\hat\rho_{n>1}$ corresponding to zero, one, and more than one photon, occurring with probabilities $P_0$, $P_1$, and $P_{n>1}$, respectively. For a good single photon source we can then relate the visibilities and the second order correlation function through the relation 
\begin{equation}
  V=V_1(1-F g^{(2)})\,.
   \label{eq:F_factor}
\end{equation}
where $V_1$ is the visibility of the single photon component, $F$ is a factor of order unity, and we have neglected higher order terms in $g^{(2)}$ and $1-V_1$ since these are both expected to be small for a good single photon source. 
A factor of $F=2$ is commonly employed in experiments to account for the reduction in visibility due to two-photon pollution of single-photon sources in an attempt to extract the bare HOM visibility in the absence of the two-photon component~\cite{somaschi2016near,wang2019towards}. As we show in Appendix \ref{app:flimits_reducedV}, however, $F$ depends on the degree of distinguishability between the single-photon component and the additional photons. In the simplest case of a pure single photon component $V_1=1$ with temporal mode function $\psi(t)$ and $g^{(2)}$ originating solely from the two-photon component (neglecting three or more photons), we show that 
\begin{equation}
    F=1 +P_1 \langle \hat n_{\neq\psi}\rangle_{n>1},
    \label{eq:F_fully_simplified}
\end{equation}
where  $\langle \hat n_{\neq\psi}\rangle_{n>1}$ is the number of photons in other modes than $\psi$ for the multiphoton component.

We can then consider various examples for a  single photon source with a near unity probability of producing a single photon $P_1\approx 1$. (i) If the two-photon component consists of two photons in the same mode as the single photon component $\psi$, we have $\langle \hat n_{\neq\psi}\rangle_{n>1}=0$ and $F=1$. (ii) If the two-photon component consists of a one photon in the same mode as the single photon and one distinguishable photon in a different mode, we have $\langle \hat n_{\neq\psi}\rangle_{n>1}=1$ and $F=2$. This thus corresponds to the choice often used in experiments and is equivalent to adding, e.g., dark counts in detectors with some probability and neglecting any interference between the dark counts and the single-photon component.  (iii) If both photons in the two-photon component are in modes distinguishable from the single photon, $\langle \hat n_{\neq\psi}\rangle_{n>1}=2$ and $F=3$.
Depending on the nature of the multi-photon component, $F$ can thus be in the interval $F\in [1,3]$. As we show in Appendix \ref{app:flimits_reducedV}, we cannot go outside this interval even if we allow for higher photon number components or mixed single photon states as long as $g^{(2)}\ll 1$. If we go completely away from this regime, the situation is different, and for instance we have $F=1/2$ for a coherent state with $V=1/2$ and $g^{(2)}=1$. 

\begin{figure}[htb!]
 \includegraphics[width=0.95\columnwidth]{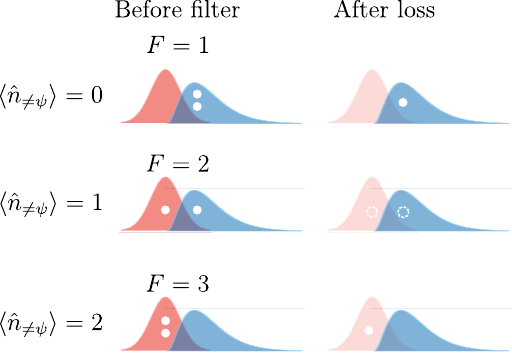}
{ \caption{Explanation of the relation between $g^{(2)}$ and indistinguishability. Photons in the waveguide can be either in the same mode as the single-photon component (blue, corresponding to the mode $\psi$ emitted by the emitter), or in a different mode (red, e.g. the laser mode). Top: If both photons are in the emitter-mode $\psi$, then the state after loss is in a pure single photon state with the correct mode function. We thus have the minimum reduction in indistinguishability, corresponding to $F=1$. Center: If one photon is in the emitter-mode $\psi$ and the other is in another mode, loss will produce a state with an equal probability of the photon to be in the right and wrong modes, leading to $F=2$. Bottom: If both photons are in the undesired mode, the photon remaining after loss will with certainty be distinguishable from the single-photon component and we have the maximum reduction in indistinguishability $F=3$.}
\label{fig:explain_F}}
\end{figure}

 To understand the origin of the factor $F$ in Eq.~\eqref{eq:F_fully_simplified} we recall that we consider the limit where there is a large probability of photon loss before reaching the detectors. In a HOM experiment detecting two-photon coincidences from two sources with non-vanishing $g^{(2)}$, there is thus always a probability that both photons originated from the same source, while the photon from the other source was lost. This yields the first contribution of unity in Eq.~\eqref{eq:F_fully_simplified}. 
The reason for the importance of the number of photons $\langle \hat n_{\neq\psi}\rangle_{n>1}$ in other modes than the single-photon mode $\psi$, can be understood by considering what happens to the two-photon component when one of the photons is lost. This is illustrated in Fig.~\ref{fig:explain_F}. If both photons were in the same mode $\psi$ as the single-photon component ($\langle \hat n_{\neq\psi}\rangle_{n>1}=0$), the state after losing a photon will be a single photon in mode $\psi$, which is the indistinguishable from the (original) single-photon component. Hence, there will be no additional reduction in $V$. On the other hand, if one of the photons was in a different mode ($\langle \hat n_{\neq\psi}\rangle_{n>1}=1$) there is a 50\% probability of ending with a photon in a different mode, resulting in a reduced visibility. If both photons were in different modes ($\langle \hat n_{\neq\psi}\rangle_{n>1}=2$), the remaining photon after loss would with certainty be in a mode different from $\psi$ and the effect would be twice as large. 

From the above discussion follows that losses change the visibility of the single photon state $V_1$, since it can be contaminated from the two-photon component by modes different from the mode $\psi$. Therefore, while our expressions for $V$ and $g^{(2)}$ are independent of the precise amount of losses before the detection, the decomposition in Eq.~\eqref{eq:F_factor} is not. This is immediately clear from Eq.~\eqref{eq:F_fully_simplified}, since the single-photon probability $P_1$ is changed by loses. Any application of Eq.~\eqref{eq:F_factor} is thus always relative to a specific decomposition point, e.g. right after the emitter. 

To gain insight into the subtle interplay between pulse length, leakage, two-photon pollution and visibility, we now quantitatively explore the relation in Eq.\eqref{eq:F_factor} for the considered single photon sources. For the unfiltered case, our simulations consider no sources of decoherence and hence the single photon density matrix is a pure single photon state with perfect visibility $V_1=1$ (note that since the leaked field acts as a displacement operator on the state, this conclusion applies even if the leaked field is non-zero $x\neq 0$). Hence, from our previous results we can evaluate $F$ from 
\begin{equation}
    F=\frac{1-V}{g^{(2)}}.
    \label{eq:F_evaluate}
\end{equation}
On the other hand, for the filtered case the photon field is in a pure state before the filter, but the filter might remove one out of two photons emitted by the emitter. Hence, the single photon component will in general not be in a pure state after the filter, so that $V_1<1$. Directly evaluating $V_1$ in our simulations is challenging since it would require evaluating even higher order correlation functions \cite{hellebek2024}. We therefore still use Eq.~\eqref{eq:F_evaluate} to evaluate the influence of the multi-photon component on the visibility $V$, since any deviation from unit visibility can only arise from multi-photon events before the filter. Since this does not exactly correspond to the situation for which the result in Eq.~\eqref{eq:F_fully_simplified} is derived, we are not necessarily bounded by $F\in [1,3]$, as we characterize the effect of the two-photon component {\it before} the filter in terms of $g^{(2)}$ {\it after} the filter, and this makes $F$ unbounded. Nevertheless, we find below that our results fall in the interval $F\in [1,3]$  except for  special cases where $1-V$ and $g^{(2)}$ almost vanish. Here the results can go slightly outside this interval in some of our simulations.

In Fig.~\ref{fig:F_factor}, the factor $F$ is plotted for two different leakage factors $|x|$ over a range of phases and pulse lengths. 
We first note that near the minimum of $g^{(2)}$, where we have a good single photon source, $F$ is in the interval $F\in[1,3]$ for the unfiltered case in accordance with the discussion above. For long and short pulses where the output in the unfiltered case approach a coherent state, $F$ approaches the value of a coherent state $F=1/2$. For $\theta=0$ the value of $F$ is strongly peaked around the minimum of $g^{(2)}$. This is a result of the destructive interference between the leaked field and the field emitted by the emitter. At the minimum of $g^{(2)}$, the component of the leaked field resembling the emission from the emitter is almost entirely eliminated by destructive interference. As a consequence, the remaining two-photon component is highly distinguishable from the single-photon component, yielding a high value of $F$. 
For the other phases, a lower value is obtained, since there is no such destructive interference and even constructive interference for $\theta=\pi$.

As seen in Fig.~\ref{fig:F_factor}, the factor $F$ is typically decreased by the filter since the filter in frequency stretches photons in time, so that photons leaked or emitted during the short pulse become more similar to the single photon pulse (thus decreasing $\langle \hat n_{\neq\psi}\rangle_{n>1}$). $F$ is, however, increased by the application of the low-pass filter in the short-pulse limit.
An explanation of this behaviour is that the filter has little influence on the photon emitted by the emitter, but largely filters out the leaked coherent field. On average there is much less than one leaked photon as evidenced by the low value of $g^{(2)}$ in the short pulse limit of Fig.~\ref{fig:g2_pulsewidth}. In the short pulse limit, the two-photon state of the filtered source thus consists of a single photon emitted by the emitter and a (somewhat) distinguishable photon from the leaked field. It thus does not approach a coherent state with $F=1/2$, as it is the case for the unfiltered field. A different exception arises for the case of destructive interference, where $1-V$ and $g^{(2)}$ almost vanish. Here, the subtle differences in the precise meaning of $F$ for the filtered and unfiltered case are important and hence the meaning of this is hard to interpret.

%******** section-5 conclusions ********
\section{\label{sec-5Conclu} Conclusions and Outlook}

We have studied the two-photon correlations of an emitter based single-photon source in the realistic scenario where the drive field may leak into the detection path. We find that the pulse length of the input field strongly influences the impurity of the single-photon source. In particular, the second-order correlation function $g^{(2)}$ reaches a minimum for a certain pulse length, which represents a compromise between minimizing the re-excitation error of the emitter and minimizing the number of photons from the leaked field. As a consequence, the optimal pulse strongly depends on the amount of laser leakage from the input field. Importantly, the phase of the leaked field matters as there will be interference with the field emitted from the emitter and one cannot simply add the two contributions. 
Moreover, we show how the leakage fraction is reduced by means of a low-pass filter, which thus strongly reduces $g^{(2)}$. 

We also study the relation between the second order correlation function and the HOM visibility $V$. Contrary to the common assumption in the literature of $F = (1 - V)/g^{(2)} = 2$, the factor $F$ can attain values anywhere in the interval between $F=1$ and $F=3$ for a good single photon source ($g^{2}\ll 1$) depending on 
the precise nature of the multiphoton component. 
The distinguishability of the generated photons will depend on the multi-photon emission of the sources, and thus the input pulse length and the phase of the leaked field. By choosing the pulse length at which $g^{(2)}$ is minimum, we generally also maximize the HOM visibility $V$, but the precise relation between $V$ and $g^{(2)}$ depends on the leakage fraction $|x|$ and phase $\theta$. Our analysis demonstrates that the standard procedure for accounting for visibility reduction due to two-photon pollution and using this to extract an intrinsic visibility is problematic and should not be used. In fact, our analysis shows that since such a decomposition depends on the point at which it is performed, it is not even clear what such an intrinsic visibility describes, e.g., a frequency filter modifies both the single- and two-photon components so that different decompositions would be obtained before and after.
Without additional measurements, i.e., a full tomography of the quantum state, it is thus not possibly to fully characterize the state.

Single-photon sources are a key asset for the development of quantum technologies. We believe that the results of this article provide a solid theoretical basis for obtaining a better understanding of their properties and thus eventually improving their capabilities. We would like to stress that our results show the importance of the scattering phase $\theta$ in addition to the leakage fraction $\abs{x}$. This should be taken into account in the future development of single photon sources, since \textit{both} leakage fraction and phase must be accounted for when optimising the input laser beam. We have introduced a possible method of determining these parameters experimentally, allowing a complete characterization of the source. This allows calculating the expected impurity and visibility, that can allow further testing and point out imperfections in upcoming experimental implementations.

\section*{Acknowledgements}
We are grateful to Ying Wang, Ming-Lai Chan, Peter Lodahl and Klaus Mølmer for useful discussions. 
Hy-Q acknowledges the support of Danmarks Grundforskningsfond (DNRF 139, Hy-Q Center for Hybrid Quantum Networks).

$^\dagger$ E.M.G.R. and J.B. contributed equally to this work.

\bibliographystyle{apsrev4-1}
\bibliography{bibs_used}% Produces the bibliography via BibTeX.

%merlin.mbs apsrev4-1.bst 2010-07-25 4.21a (PWD, AO, DPC) hacked
%Control: key (0)
%Control: author (72) initials jnrlst
%Control: editor formatted (1) identically to author
%Control: production of article title (-1) disabled
%Control: page (0) single
%Control: year (1) truncated
%Control: production of eprint (0) enabled
\begin{thebibliography}{42}%
\makeatletter
\providecommand \@ifxundefined [1]{%
 \@ifx{#1\undefined}
}%
\providecommand \@ifnum [1]{%
 \ifnum #1\expandafter \@firstoftwo
 \else \expandafter \@secondoftwo
 \fi
}%
\providecommand \@ifx [1]{%
 \ifx #1\expandafter \@firstoftwo
 \else \expandafter \@secondoftwo
 \fi
}%
\providecommand \natexlab [1]{#1}%
\providecommand \enquote  [1]{``#1''}%
\providecommand \bibnamefont  [1]{#1}%
\providecommand \bibfnamefont [1]{#1}%
\providecommand \citenamefont [1]{#1}%
\providecommand \href@noop [0]{\@secondoftwo}%
\providecommand \href [0]{\begingroup \@sanitize@url \@href}%
\providecommand \@href[1]{\@@startlink{#1}\@@href}%
\providecommand \@@href[1]{\endgroup#1\@@endlink}%
\providecommand \@sanitize@url [0]{\catcode `\\12\catcode `\$12\catcode `\&12\catcode `\#12\catcode `\^12\catcode `\_12\catcode `\%12\relax}%
\providecommand \@@startlink[1]{}%
\providecommand \@@endlink[0]{}%
\providecommand \url  [0]{\begingroup\@sanitize@url \@url }%
\providecommand \@url [1]{\endgroup\@href {#1}{\urlprefix }}%
\providecommand \urlprefix  [0]{URL }%
\providecommand \Eprint [0]{\href }%
\providecommand \doibase [0]{http://dx.doi.org/}%
\providecommand \selectlanguage [0]{\@gobble}%
\providecommand \bibinfo  [0]{\@secondoftwo}%
\providecommand \bibfield  [0]{\@secondoftwo}%
\providecommand \translation [1]{[#1]}%
\providecommand \BibitemOpen [0]{}%
\providecommand \bibitemStop [0]{}%
\providecommand \bibitemNoStop [0]{.\EOS\space}%
\providecommand \EOS [0]{\spacefactor3000\relax}%
\providecommand \BibitemShut  [1]{\csname bibitem#1\endcsname}%
\let\auto@bib@innerbib\@empty
%</preamble>
\bibitem [{\citenamefont {Thomas}\ and\ \citenamefont {Senellart}(2021)}]{thomas2021race}%
  \BibitemOpen
  \bibfield  {author} {\bibinfo {author} {\bibfnamefont {S.}~\bibnamefont {Thomas}}\ and\ \bibinfo {author} {\bibfnamefont {P.}~\bibnamefont {Senellart}},\ }\href@noop {} {\bibfield  {journal} {\bibinfo  {journal} {Nature Nanotechnology}\ }\textbf {\bibinfo {volume} {16}},\ \bibinfo {pages} {367} (\bibinfo {year} {2021})}\BibitemShut {NoStop}%
\bibitem [{\citenamefont {Knill}\ \emph {et~al.}(2001)\citenamefont {Knill}, \citenamefont {Laflamme},\ and\ \citenamefont {Milburn}}]{knill2001scheme}%
  \BibitemOpen
  \bibfield  {author} {\bibinfo {author} {\bibfnamefont {E.}~\bibnamefont {Knill}}, \bibinfo {author} {\bibfnamefont {R.}~\bibnamefont {Laflamme}}, \ and\ \bibinfo {author} {\bibfnamefont {G.~J.}\ \bibnamefont {Milburn}},\ }\href@noop {} {\bibfield  {journal} {\bibinfo  {journal} {nature}\ }\textbf {\bibinfo {volume} {409}},\ \bibinfo {pages} {46} (\bibinfo {year} {2001})}\BibitemShut {NoStop}%
\bibitem [{\citenamefont {O'brien}(2007)}]{o2007optical}%
  \BibitemOpen
  \bibfield  {author} {\bibinfo {author} {\bibfnamefont {J.~L.}\ \bibnamefont {O'brien}},\ }\href@noop {} {\bibfield  {journal} {\bibinfo  {journal} {Science}\ }\textbf {\bibinfo {volume} {318}},\ \bibinfo {pages} {1567} (\bibinfo {year} {2007})}\BibitemShut {NoStop}%
\bibitem [{\citenamefont {{Aspuru-Guzik}}\ and\ \citenamefont {Walther}(2012)}]{aspuru2012photonic}%
  \BibitemOpen
  \bibfield  {author} {\bibinfo {author} {\bibfnamefont {A.}~\bibnamefont {{Aspuru-Guzik}}}\ and\ \bibinfo {author} {\bibfnamefont {P.}~\bibnamefont {Walther}},\ }\href@noop {} {\bibfield  {journal} {\bibinfo  {journal} {Nature physics}\ }\textbf {\bibinfo {volume} {8}},\ \bibinfo {pages} {285} (\bibinfo {year} {2012})}\BibitemShut {NoStop}%
\bibitem [{\citenamefont {Duan}\ \emph {et~al.}(2001)\citenamefont {Duan}, \citenamefont {Lukin}, \citenamefont {Cirac},\ and\ \citenamefont {Zoller}}]{duan2001long}%
  \BibitemOpen
  \bibfield  {author} {\bibinfo {author} {\bibfnamefont {L.-M.}\ \bibnamefont {Duan}}, \bibinfo {author} {\bibfnamefont {M.~D.}\ \bibnamefont {Lukin}}, \bibinfo {author} {\bibfnamefont {J.~I.}\ \bibnamefont {Cirac}}, \ and\ \bibinfo {author} {\bibfnamefont {P.}~\bibnamefont {Zoller}},\ }\href@noop {} {\bibfield  {journal} {\bibinfo  {journal} {Nature}\ }\textbf {\bibinfo {volume} {414}},\ \bibinfo {pages} {413} (\bibinfo {year} {2001})}\BibitemShut {NoStop}%
\bibitem [{\citenamefont {Sangouard}\ \emph {et~al.}(2011)\citenamefont {Sangouard}, \citenamefont {Simon}, \citenamefont {De~Riedmatten},\ and\ \citenamefont {Gisin}}]{sangouard2011quantum}%
  \BibitemOpen
  \bibfield  {author} {\bibinfo {author} {\bibfnamefont {N.}~\bibnamefont {Sangouard}}, \bibinfo {author} {\bibfnamefont {C.}~\bibnamefont {Simon}}, \bibinfo {author} {\bibfnamefont {H.}~\bibnamefont {De~Riedmatten}}, \ and\ \bibinfo {author} {\bibfnamefont {N.}~\bibnamefont {Gisin}},\ }\href@noop {} {\bibfield  {journal} {\bibinfo  {journal} {Reviews of Modern Physics}\ }\textbf {\bibinfo {volume} {83}},\ \bibinfo {pages} {33} (\bibinfo {year} {2011})}\BibitemShut {NoStop}%
\bibitem [{\citenamefont {Kimble}(2008)}]{kimble2008quantum}%
  \BibitemOpen
  \bibfield  {author} {\bibinfo {author} {\bibfnamefont {H.~J.}\ \bibnamefont {Kimble}},\ }\href@noop {} {\bibfield  {journal} {\bibinfo  {journal} {Nature}\ }\textbf {\bibinfo {volume} {453}},\ \bibinfo {pages} {1023} (\bibinfo {year} {2008})}\BibitemShut {NoStop}%
\bibitem [{\citenamefont {Tomm}\ \emph {et~al.}(2021)\citenamefont {Tomm}, \citenamefont {Javadi}, \citenamefont {Antoniadis}, \citenamefont {Najer}, \citenamefont {L{\"o}bl}, \citenamefont {Korsch}, \citenamefont {Schott}, \citenamefont {Valentin}, \citenamefont {Wieck}, \citenamefont {Ludwig},\ and\ \citenamefont {Warburton}}]{tomm_bright_2021}%
  \BibitemOpen
  \bibfield  {author} {\bibinfo {author} {\bibfnamefont {N.}~\bibnamefont {Tomm}}, \bibinfo {author} {\bibfnamefont {A.}~\bibnamefont {Javadi}}, \bibinfo {author} {\bibfnamefont {N.~O.}\ \bibnamefont {Antoniadis}}, \bibinfo {author} {\bibfnamefont {D.}~\bibnamefont {Najer}}, \bibinfo {author} {\bibfnamefont {M.~C.}\ \bibnamefont {L{\"o}bl}}, \bibinfo {author} {\bibfnamefont {A.~R.}\ \bibnamefont {Korsch}}, \bibinfo {author} {\bibfnamefont {R.}~\bibnamefont {Schott}}, \bibinfo {author} {\bibfnamefont {S.~R.}\ \bibnamefont {Valentin}}, \bibinfo {author} {\bibfnamefont {A.~D.}\ \bibnamefont {Wieck}}, \bibinfo {author} {\bibfnamefont {A.}~\bibnamefont {Ludwig}}, \ and\ \bibinfo {author} {\bibfnamefont {R.~J.}\ \bibnamefont {Warburton}},\ }\href {\doibase 10.1038/s41565-020-00831-x} {\bibfield  {journal} {\bibinfo  {journal} {Nature Nanotechnology}\ }\textbf {\bibinfo {volume} {16}},\ \bibinfo {pages} {399} (\bibinfo {year} {2021})}\BibitemShut {NoStop}%
\bibitem [{\citenamefont {Aharonovich}\ \emph {et~al.}(2016)\citenamefont {Aharonovich}, \citenamefont {Englund},\ and\ \citenamefont {Toth}}]{aharonovich2016solid}%
  \BibitemOpen
  \bibfield  {author} {\bibinfo {author} {\bibfnamefont {I.}~\bibnamefont {Aharonovich}}, \bibinfo {author} {\bibfnamefont {D.}~\bibnamefont {Englund}}, \ and\ \bibinfo {author} {\bibfnamefont {M.}~\bibnamefont {Toth}},\ }\href@noop {} {\bibfield  {journal} {\bibinfo  {journal} {Nature Photonics}\ }\textbf {\bibinfo {volume} {10}},\ \bibinfo {pages} {631} (\bibinfo {year} {2016})}\BibitemShut {NoStop}%
\bibitem [{\citenamefont {Senellart}\ \emph {et~al.}(2017)\citenamefont {Senellart}, \citenamefont {Solomon},\ and\ \citenamefont {White}}]{senellart2017high}%
  \BibitemOpen
  \bibfield  {author} {\bibinfo {author} {\bibfnamefont {P.}~\bibnamefont {Senellart}}, \bibinfo {author} {\bibfnamefont {G.}~\bibnamefont {Solomon}}, \ and\ \bibinfo {author} {\bibfnamefont {A.}~\bibnamefont {White}},\ }\href@noop {} {\bibfield  {journal} {\bibinfo  {journal} {Nature nanotechnology}\ }\textbf {\bibinfo {volume} {12}},\ \bibinfo {pages} {1026} (\bibinfo {year} {2017})}\BibitemShut {NoStop}%
\bibitem [{\citenamefont {Wang}\ \emph {et~al.}(2016)\citenamefont {Wang}, \citenamefont {Duan}, \citenamefont {Li}, \citenamefont {Chen}, \citenamefont {Li}, \citenamefont {He}, \citenamefont {Chen}, \citenamefont {He}, \citenamefont {Ding}, \citenamefont {Peng} \emph {et~al.}}]{wang2016near}%
  \BibitemOpen
  \bibfield  {author} {\bibinfo {author} {\bibfnamefont {H.}~\bibnamefont {Wang}}, \bibinfo {author} {\bibfnamefont {Z.-C.}\ \bibnamefont {Duan}}, \bibinfo {author} {\bibfnamefont {Y.-H.}\ \bibnamefont {Li}}, \bibinfo {author} {\bibfnamefont {S.}~\bibnamefont {Chen}}, \bibinfo {author} {\bibfnamefont {J.-P.}\ \bibnamefont {Li}}, \bibinfo {author} {\bibfnamefont {Y.-M.}\ \bibnamefont {He}}, \bibinfo {author} {\bibfnamefont {M.-C.}\ \bibnamefont {Chen}}, \bibinfo {author} {\bibfnamefont {Y.}~\bibnamefont {He}}, \bibinfo {author} {\bibfnamefont {X.}~\bibnamefont {Ding}}, \bibinfo {author} {\bibfnamefont {C.-Z.}\ \bibnamefont {Peng}},  \emph {et~al.},\ }\href@noop {} {\bibfield  {journal} {\bibinfo  {journal} {Physical Review Letters}\ }\textbf {\bibinfo {volume} {116}},\ \bibinfo {pages} {213601} (\bibinfo {year} {2016})}\BibitemShut {NoStop}%
\bibitem [{\citenamefont {Arakawa}\ and\ \citenamefont {Holmes}(2020)}]{arakawa_progress_2020}%
  \BibitemOpen
  \bibfield  {author} {\bibinfo {author} {\bibfnamefont {Y.}~\bibnamefont {Arakawa}}\ and\ \bibinfo {author} {\bibfnamefont {M.~J.}\ \bibnamefont {Holmes}},\ }\href {\doibase 10.1063/5.0010193} {\bibfield  {journal} {\bibinfo  {journal} {Applied Physics Reviews}\ }\textbf {\bibinfo {volume} {7}},\ \bibinfo {pages} {021309} (\bibinfo {year} {2020})}\BibitemShut {NoStop}%
\bibitem [{\citenamefont {Ollivier}\ \emph {et~al.}(2020)\citenamefont {Ollivier}, \citenamefont {{Maillette de Buy Wenniger}}, \citenamefont {Thomas}, \citenamefont {Wein}, \citenamefont {Harouri}, \citenamefont {Coppola}, \citenamefont {Hilaire}, \citenamefont {Millet}, \citenamefont {Lema{\^i}tre}, \citenamefont {Sagnes}, \citenamefont {Krebs}, \citenamefont {Lanco}, \citenamefont {Loredo}, \citenamefont {Ant{\'o}n}, \citenamefont {Somaschi},\ and\ \citenamefont {Senellart}}]{ollivier_reproducibility_2020}%
  \BibitemOpen
  \bibfield  {author} {\bibinfo {author} {\bibfnamefont {H.}~\bibnamefont {Ollivier}}, \bibinfo {author} {\bibfnamefont {I.}~\bibnamefont {{Maillette de Buy Wenniger}}}, \bibinfo {author} {\bibfnamefont {S.}~\bibnamefont {Thomas}}, \bibinfo {author} {\bibfnamefont {S.~C.}\ \bibnamefont {Wein}}, \bibinfo {author} {\bibfnamefont {A.}~\bibnamefont {Harouri}}, \bibinfo {author} {\bibfnamefont {G.}~\bibnamefont {Coppola}}, \bibinfo {author} {\bibfnamefont {P.}~\bibnamefont {Hilaire}}, \bibinfo {author} {\bibfnamefont {C.}~\bibnamefont {Millet}}, \bibinfo {author} {\bibfnamefont {A.}~\bibnamefont {Lema{\^i}tre}}, \bibinfo {author} {\bibfnamefont {I.}~\bibnamefont {Sagnes}}, \bibinfo {author} {\bibfnamefont {O.}~\bibnamefont {Krebs}}, \bibinfo {author} {\bibfnamefont {L.}~\bibnamefont {Lanco}}, \bibinfo {author} {\bibfnamefont {J.~C.}\ \bibnamefont {Loredo}}, \bibinfo {author} {\bibfnamefont {C.}~\bibnamefont {Ant{\'o}n}}, \bibinfo {author} {\bibfnamefont {N.}~\bibnamefont {Somaschi}}, \ and\ \bibinfo {author}
  {\bibfnamefont {P.}~\bibnamefont {Senellart}},\ }\href {\doibase 10.1021/acsphotonics.9b01805} {\bibfield  {journal} {\bibinfo  {journal} {ACS Photonics}\ }\textbf {\bibinfo {volume} {7}},\ \bibinfo {pages} {1050} (\bibinfo {year} {2020})}\BibitemShut {NoStop}%
\bibitem [{\citenamefont {Laferri{\`e}re}\ \emph {et~al.}(2022)\citenamefont {Laferri{\`e}re}, \citenamefont {Yeung}, \citenamefont {Miron}, \citenamefont {Northeast}, \citenamefont {Haffouz}, \citenamefont {Lapointe}, \citenamefont {Korkusinski}, \citenamefont {Poole}, \citenamefont {Williams},\ and\ \citenamefont {Dalacu}}]{laferriere_unity_2022}%
  \BibitemOpen
  \bibfield  {author} {\bibinfo {author} {\bibfnamefont {P.}~\bibnamefont {Laferri{\`e}re}}, \bibinfo {author} {\bibfnamefont {E.}~\bibnamefont {Yeung}}, \bibinfo {author} {\bibfnamefont {I.}~\bibnamefont {Miron}}, \bibinfo {author} {\bibfnamefont {D.~B.}\ \bibnamefont {Northeast}}, \bibinfo {author} {\bibfnamefont {S.}~\bibnamefont {Haffouz}}, \bibinfo {author} {\bibfnamefont {J.}~\bibnamefont {Lapointe}}, \bibinfo {author} {\bibfnamefont {M.}~\bibnamefont {Korkusinski}}, \bibinfo {author} {\bibfnamefont {P.~J.}\ \bibnamefont {Poole}}, \bibinfo {author} {\bibfnamefont {R.~L.}\ \bibnamefont {Williams}}, \ and\ \bibinfo {author} {\bibfnamefont {D.}~\bibnamefont {Dalacu}},\ }\href {\doibase 10.1038/s41598-022-10451-1} {\bibfield  {journal} {\bibinfo  {journal} {Scientific Reports}\ }\textbf {\bibinfo {volume} {12}},\ \bibinfo {pages} {6376} (\bibinfo {year} {2022})}\BibitemShut {NoStop}%
\bibitem [{\citenamefont {Lu}\ and\ \citenamefont {Pan}(2021)}]{lu_quantumdot_2021}%
  \BibitemOpen
  \bibfield  {author} {\bibinfo {author} {\bibfnamefont {C.-Y.}\ \bibnamefont {Lu}}\ and\ \bibinfo {author} {\bibfnamefont {J.-W.}\ \bibnamefont {Pan}},\ }\href {\doibase 10.1038/s41565-021-01033-9} {\bibfield  {journal} {\bibinfo  {journal} {Nature Nanotechnology}\ }\textbf {\bibinfo {volume} {16}},\ \bibinfo {pages} {1294} (\bibinfo {year} {2021})}\BibitemShut {NoStop}%
\bibitem [{\citenamefont {Valeri}\ \emph {et~al.}(2024)\citenamefont {Valeri}, \citenamefont {Barigelli}, \citenamefont {Polacchi}, \citenamefont {Rodari}, \citenamefont {Santis}, \citenamefont {Giordani}, \citenamefont {Carvacho}, \citenamefont {Spagnolo},\ and\ \citenamefont {Sciarrino}}]{valeri_generation_2024}%
  \BibitemOpen
  \bibfield  {author} {\bibinfo {author} {\bibfnamefont {M.}~\bibnamefont {Valeri}}, \bibinfo {author} {\bibfnamefont {P.}~\bibnamefont {Barigelli}}, \bibinfo {author} {\bibfnamefont {B.}~\bibnamefont {Polacchi}}, \bibinfo {author} {\bibfnamefont {G.}~\bibnamefont {Rodari}}, \bibinfo {author} {\bibfnamefont {G.~D.}\ \bibnamefont {Santis}}, \bibinfo {author} {\bibfnamefont {T.}~\bibnamefont {Giordani}}, \bibinfo {author} {\bibfnamefont {G.}~\bibnamefont {Carvacho}}, \bibinfo {author} {\bibfnamefont {N.}~\bibnamefont {Spagnolo}}, \ and\ \bibinfo {author} {\bibfnamefont {F.}~\bibnamefont {Sciarrino}},\ }\href {\doibase 10.1088/2058-9565/ad1c44} {\bibfield  {journal} {\bibinfo  {journal} {Quantum Science and Technology}\ }\textbf {\bibinfo {volume} {9}},\ \bibinfo {pages} {025002} (\bibinfo {year} {2024})}\BibitemShut {NoStop}%
\bibitem [{\citenamefont {Uppu}\ \emph {et~al.}(2020)\citenamefont {Uppu}, \citenamefont {Pedersen}, \citenamefont {Wang}, \citenamefont {Olesen}, \citenamefont {Papon}, \citenamefont {Zhou}, \citenamefont {Midolo}, \citenamefont {Scholz}, \citenamefont {Wieck}, \citenamefont {Ludwig} \emph {et~al.}}]{uppu2020scalable}%
  \BibitemOpen
  \bibfield  {author} {\bibinfo {author} {\bibfnamefont {R.}~\bibnamefont {Uppu}}, \bibinfo {author} {\bibfnamefont {F.~T.}\ \bibnamefont {Pedersen}}, \bibinfo {author} {\bibfnamefont {Y.}~\bibnamefont {Wang}}, \bibinfo {author} {\bibfnamefont {C.~T.}\ \bibnamefont {Olesen}}, \bibinfo {author} {\bibfnamefont {C.}~\bibnamefont {Papon}}, \bibinfo {author} {\bibfnamefont {X.}~\bibnamefont {Zhou}}, \bibinfo {author} {\bibfnamefont {L.}~\bibnamefont {Midolo}}, \bibinfo {author} {\bibfnamefont {S.}~\bibnamefont {Scholz}}, \bibinfo {author} {\bibfnamefont {A.~D.}\ \bibnamefont {Wieck}}, \bibinfo {author} {\bibfnamefont {A.}~\bibnamefont {Ludwig}},  \emph {et~al.},\ }\href@noop {} {\bibfield  {journal} {\bibinfo  {journal} {Science advances}\ }\textbf {\bibinfo {volume} {6}},\ \bibinfo {pages} {eabc8268} (\bibinfo {year} {2020})}\BibitemShut {NoStop}%
\bibitem [{\citenamefont {{Gonz{\'a}lez-Ruiz}}\ \emph {et~al.}(2022{\natexlab{a}})\citenamefont {{Gonz{\'a}lez-Ruiz}}, \citenamefont {Das}, \citenamefont {Lodahl},\ and\ \citenamefont {S{\o}rensen}}]{gonzalez2022violation}%
  \BibitemOpen
  \bibfield  {author} {\bibinfo {author} {\bibfnamefont {E.~M.}\ \bibnamefont {{Gonz{\'a}lez-Ruiz}}}, \bibinfo {author} {\bibfnamefont {S.~K.}\ \bibnamefont {Das}}, \bibinfo {author} {\bibfnamefont {P.}~\bibnamefont {Lodahl}}, \ and\ \bibinfo {author} {\bibfnamefont {A.~S.}\ \bibnamefont {S{\o}rensen}},\ }\href@noop {} {\bibfield  {journal} {\bibinfo  {journal} {Physical Review A}\ }\textbf {\bibinfo {volume} {106}},\ \bibinfo {pages} {012222} (\bibinfo {year} {2022}{\natexlab{a}})}\BibitemShut {NoStop}%
\bibitem [{\citenamefont {{Gonz{\'a}lez-Ruiz}}\ \emph {et~al.}(2022{\natexlab{b}})\citenamefont {{Gonz{\'a}lez-Ruiz}}, \citenamefont {{Rivera-Dean}}, \citenamefont {Cenni}, \citenamefont {S{\o}rensen}, \citenamefont {Ac{\'i}n},\ and\ \citenamefont {Oudot}}]{gonzalez2022diqkd}%
  \BibitemOpen
  \bibfield  {author} {\bibinfo {author} {\bibfnamefont {E.~M.}\ \bibnamefont {{Gonz{\'a}lez-Ruiz}}}, \bibinfo {author} {\bibfnamefont {J.}~\bibnamefont {{Rivera-Dean}}}, \bibinfo {author} {\bibfnamefont {M.~F.~B.}\ \bibnamefont {Cenni}}, \bibinfo {author} {\bibfnamefont {A.~S.}\ \bibnamefont {S{\o}rensen}}, \bibinfo {author} {\bibfnamefont {A.}~\bibnamefont {Ac{\'i}n}}, \ and\ \bibinfo {author} {\bibfnamefont {E.}~\bibnamefont {Oudot}},\ }\href {\doibase 10.48550/arXiv.2211.16472} {\enquote {\bibinfo {title} {Device {{Independent Quantum Key Distribution}} with realistic single-photon source implementations},}\ } (\bibinfo {year} {2022}{\natexlab{b}})\BibitemShut {NoStop}%
\bibitem [{\citenamefont {Maring}\ \emph {et~al.}(2024)\citenamefont {Maring}, \citenamefont {Fyrillas}, \citenamefont {Pont}, \citenamefont {Ivanov}, \citenamefont {Stepanov}, \citenamefont {Margaria}, \citenamefont {Hease}, \citenamefont {Pishchagin}, \citenamefont {Lema{\^i}tre}, \citenamefont {Sagnes}, \citenamefont {Au}, \citenamefont {Boissier}, \citenamefont {Bertasi}, \citenamefont {Baert}, \citenamefont {Valdivia}, \citenamefont {Billard}, \citenamefont {Acar}, \citenamefont {Brieussel}, \citenamefont {Mezher}, \citenamefont {Wein}, \citenamefont {Salavrakos}, \citenamefont {Sinnott}, \citenamefont {Fioretto}, \citenamefont {Emeriau}, \citenamefont {Belabas}, \citenamefont {Mansfield}, \citenamefont {Senellart}, \citenamefont {Senellart},\ and\ \citenamefont {Somaschi}}]{maring_versatile_2024}%
  \BibitemOpen
  \bibfield  {author} {\bibinfo {author} {\bibfnamefont {N.}~\bibnamefont {Maring}}, \bibinfo {author} {\bibfnamefont {A.}~\bibnamefont {Fyrillas}}, \bibinfo {author} {\bibfnamefont {M.}~\bibnamefont {Pont}}, \bibinfo {author} {\bibfnamefont {E.}~\bibnamefont {Ivanov}}, \bibinfo {author} {\bibfnamefont {P.}~\bibnamefont {Stepanov}}, \bibinfo {author} {\bibfnamefont {N.}~\bibnamefont {Margaria}}, \bibinfo {author} {\bibfnamefont {W.}~\bibnamefont {Hease}}, \bibinfo {author} {\bibfnamefont {A.}~\bibnamefont {Pishchagin}}, \bibinfo {author} {\bibfnamefont {A.}~\bibnamefont {Lema{\^i}tre}}, \bibinfo {author} {\bibfnamefont {I.}~\bibnamefont {Sagnes}}, \bibinfo {author} {\bibfnamefont {T.~H.}\ \bibnamefont {Au}}, \bibinfo {author} {\bibfnamefont {S.}~\bibnamefont {Boissier}}, \bibinfo {author} {\bibfnamefont {E.}~\bibnamefont {Bertasi}}, \bibinfo {author} {\bibfnamefont {A.}~\bibnamefont {Baert}}, \bibinfo {author} {\bibfnamefont {M.}~\bibnamefont {Valdivia}}, \bibinfo {author} {\bibfnamefont {M.}~\bibnamefont
  {Billard}}, \bibinfo {author} {\bibfnamefont {O.}~\bibnamefont {Acar}}, \bibinfo {author} {\bibfnamefont {A.}~\bibnamefont {Brieussel}}, \bibinfo {author} {\bibfnamefont {R.}~\bibnamefont {Mezher}}, \bibinfo {author} {\bibfnamefont {S.~C.}\ \bibnamefont {Wein}}, \bibinfo {author} {\bibfnamefont {A.}~\bibnamefont {Salavrakos}}, \bibinfo {author} {\bibfnamefont {P.}~\bibnamefont {Sinnott}}, \bibinfo {author} {\bibfnamefont {D.~A.}\ \bibnamefont {Fioretto}}, \bibinfo {author} {\bibfnamefont {P.-E.}\ \bibnamefont {Emeriau}}, \bibinfo {author} {\bibfnamefont {N.}~\bibnamefont {Belabas}}, \bibinfo {author} {\bibfnamefont {S.}~\bibnamefont {Mansfield}}, \bibinfo {author} {\bibfnamefont {P.}~\bibnamefont {Senellart}}, \bibinfo {author} {\bibfnamefont {J.}~\bibnamefont {Senellart}}, \ and\ \bibinfo {author} {\bibfnamefont {N.}~\bibnamefont {Somaschi}},\ }\href {\doibase https://doi.org/10.1038/s41566-024-01403-4} {\bibfield  {journal} {\bibinfo  {journal} {Nature Photonics}\ }\textbf {\bibinfo {volume} {18}},\
  \bibinfo {pages} {603} (\bibinfo {year} {2024})}\BibitemShut {NoStop}%
\bibitem [{\citenamefont {Hong}\ \emph {et~al.}(1987)\citenamefont {Hong}, \citenamefont {Ou},\ and\ \citenamefont {Mandel}}]{hong1987measurement}%
  \BibitemOpen
  \bibfield  {author} {\bibinfo {author} {\bibfnamefont {C.-K.}\ \bibnamefont {Hong}}, \bibinfo {author} {\bibfnamefont {Z.-Y.}\ \bibnamefont {Ou}}, \ and\ \bibinfo {author} {\bibfnamefont {L.}~\bibnamefont {Mandel}},\ }\href@noop {} {\bibfield  {journal} {\bibinfo  {journal} {Physical review letters}\ }\textbf {\bibinfo {volume} {59}},\ \bibinfo {pages} {2044} (\bibinfo {year} {1987})}\BibitemShut {NoStop}%
\bibitem [{\citenamefont {Somaschi}\ \emph {et~al.}(2016)\citenamefont {Somaschi}, \citenamefont {Giesz}, \citenamefont {De~Santis}, \citenamefont {Loredo}, \citenamefont {Almeida}, \citenamefont {Hornecker}, \citenamefont {Portalupi}, \citenamefont {Grange}, \citenamefont {Anton}, \citenamefont {Demory} \emph {et~al.}}]{somaschi2016near}%
  \BibitemOpen
  \bibfield  {author} {\bibinfo {author} {\bibfnamefont {N.}~\bibnamefont {Somaschi}}, \bibinfo {author} {\bibfnamefont {V.}~\bibnamefont {Giesz}}, \bibinfo {author} {\bibfnamefont {L.}~\bibnamefont {De~Santis}}, \bibinfo {author} {\bibfnamefont {{\relax JC}.}~\bibnamefont {Loredo}}, \bibinfo {author} {\bibfnamefont {M.~P.}\ \bibnamefont {Almeida}}, \bibinfo {author} {\bibfnamefont {G.}~\bibnamefont {Hornecker}}, \bibinfo {author} {\bibfnamefont {S.~L.}\ \bibnamefont {Portalupi}}, \bibinfo {author} {\bibfnamefont {T.}~\bibnamefont {Grange}}, \bibinfo {author} {\bibfnamefont {C.}~\bibnamefont {Anton}}, \bibinfo {author} {\bibfnamefont {J.}~\bibnamefont {Demory}},  \emph {et~al.},\ }\href@noop {} {\bibfield  {journal} {\bibinfo  {journal} {Nature Photonics}\ }\textbf {\bibinfo {volume} {10}},\ \bibinfo {pages} {340} (\bibinfo {year} {2016})}\BibitemShut {NoStop}%
\bibitem [{\citenamefont {Wang}\ \emph {et~al.}(2019)\citenamefont {Wang}, \citenamefont {He}, \citenamefont {Chung}, \citenamefont {Hu}, \citenamefont {Yu}, \citenamefont {Chen}, \citenamefont {Ding}, \citenamefont {Chen}, \citenamefont {Qin}, \citenamefont {Yang} \emph {et~al.}}]{wang2019towards}%
  \BibitemOpen
  \bibfield  {author} {\bibinfo {author} {\bibfnamefont {H.}~\bibnamefont {Wang}}, \bibinfo {author} {\bibfnamefont {Y.-M.}\ \bibnamefont {He}}, \bibinfo {author} {\bibfnamefont {T.-H.}\ \bibnamefont {Chung}}, \bibinfo {author} {\bibfnamefont {H.}~\bibnamefont {Hu}}, \bibinfo {author} {\bibfnamefont {Y.}~\bibnamefont {Yu}}, \bibinfo {author} {\bibfnamefont {S.}~\bibnamefont {Chen}}, \bibinfo {author} {\bibfnamefont {X.}~\bibnamefont {Ding}}, \bibinfo {author} {\bibfnamefont {M.-C.}\ \bibnamefont {Chen}}, \bibinfo {author} {\bibfnamefont {J.}~\bibnamefont {Qin}}, \bibinfo {author} {\bibfnamefont {X.}~\bibnamefont {Yang}},  \emph {et~al.},\ }\href@noop {} {\bibfield  {journal} {\bibinfo  {journal} {Nature Photonics}\ }\textbf {\bibinfo {volume} {13}},\ \bibinfo {pages} {770} (\bibinfo {year} {2019})}\BibitemShut {NoStop}%
\bibitem [{\citenamefont {Ollivier}\ \emph {et~al.}(2021)\citenamefont {Ollivier}, \citenamefont {Thomas}, \citenamefont {Wein}, \citenamefont {{de Buy Wenniger}}, \citenamefont {Coste}, \citenamefont {Loredo}, \citenamefont {Somaschi}, \citenamefont {Harouri}, \citenamefont {Lemaitre}, \citenamefont {Sagnes} \emph {et~al.}}]{ollivier2021hong}%
  \BibitemOpen
  \bibfield  {author} {\bibinfo {author} {\bibfnamefont {H.}~\bibnamefont {Ollivier}}, \bibinfo {author} {\bibfnamefont {{\relax SE}.}~\bibnamefont {Thomas}}, \bibinfo {author} {\bibfnamefont {{\relax SC}.}~\bibnamefont {Wein}}, \bibinfo {author} {\bibfnamefont {I.~M.}\ \bibnamefont {{de Buy Wenniger}}}, \bibinfo {author} {\bibfnamefont {N.}~\bibnamefont {Coste}}, \bibinfo {author} {\bibfnamefont {{\relax JC}.}~\bibnamefont {Loredo}}, \bibinfo {author} {\bibfnamefont {N.}~\bibnamefont {Somaschi}}, \bibinfo {author} {\bibfnamefont {A.}~\bibnamefont {Harouri}}, \bibinfo {author} {\bibfnamefont {A.}~\bibnamefont {Lemaitre}}, \bibinfo {author} {\bibfnamefont {I.}~\bibnamefont {Sagnes}},  \emph {et~al.},\ }\href@noop {} {\bibfield  {journal} {\bibinfo  {journal} {Physical Review Letters}\ }\textbf {\bibinfo {volume} {126}},\ \bibinfo {pages} {063602} (\bibinfo {year} {2021})}\BibitemShut {NoStop}%
\bibitem [{\citenamefont {Loredo}\ \emph {et~al.}(2019)\citenamefont {Loredo}, \citenamefont {Antón}, \citenamefont {Reznychenko}, \citenamefont {Hilaire}, \citenamefont {Harouri}, \citenamefont {Millet}, \citenamefont {Ollivier}, \citenamefont {Somaschi}, \citenamefont {De~Santis}, \citenamefont {Lemaître}, \citenamefont {Sagnes}, \citenamefont {Lanco}, \citenamefont {Auffèves}, \citenamefont {Krebs},\ and\ \citenamefont {Senellart}}]{loredo_generation_2019}%
  \BibitemOpen
  \bibfield  {author} {\bibinfo {author} {\bibfnamefont {J.~C.}\ \bibnamefont {Loredo}}, \bibinfo {author} {\bibfnamefont {C.}~\bibnamefont {Antón}}, \bibinfo {author} {\bibfnamefont {B.}~\bibnamefont {Reznychenko}}, \bibinfo {author} {\bibfnamefont {P.}~\bibnamefont {Hilaire}}, \bibinfo {author} {\bibfnamefont {A.}~\bibnamefont {Harouri}}, \bibinfo {author} {\bibfnamefont {C.}~\bibnamefont {Millet}}, \bibinfo {author} {\bibfnamefont {H.}~\bibnamefont {Ollivier}}, \bibinfo {author} {\bibfnamefont {N.}~\bibnamefont {Somaschi}}, \bibinfo {author} {\bibfnamefont {L.}~\bibnamefont {De~Santis}}, \bibinfo {author} {\bibfnamefont {A.}~\bibnamefont {Lemaître}}, \bibinfo {author} {\bibfnamefont {I.}~\bibnamefont {Sagnes}}, \bibinfo {author} {\bibfnamefont {L.}~\bibnamefont {Lanco}}, \bibinfo {author} {\bibfnamefont {A.}~\bibnamefont {Auffèves}}, \bibinfo {author} {\bibfnamefont {O.}~\bibnamefont {Krebs}}, \ and\ \bibinfo {author} {\bibfnamefont {P.}~\bibnamefont {Senellart}},\ }\href {\doibase
  10.1038/s41566-019-0506-3} {\bibfield  {journal} {\bibinfo  {journal} {Nature Photonics}\ }\textbf {\bibinfo {volume} {13}},\ \bibinfo {pages} {803} (\bibinfo {year} {2019})},\ \bibinfo {note} {publisher: Nature Publishing Group}\BibitemShut {NoStop}%
\bibitem [{\citenamefont {Wenniger}\ \emph {et~al.}(2024)\citenamefont {Wenniger}, \citenamefont {Wein}, \citenamefont {Fioretto}, \citenamefont {Thomas}, \citenamefont {Antón-Solanas}, \citenamefont {Lemaître}, \citenamefont {Sagnes}, \citenamefont {Harouri}, \citenamefont {Belabas}, \citenamefont {Somaschi}, \citenamefont {Hilaire}, \citenamefont {Senellart},\ and\ \citenamefont {Senellart}}]{wenniger_quantum_2024}%
  \BibitemOpen
  \bibfield  {author} {\bibinfo {author} {\bibfnamefont {I.~M. d.~B.}\ \bibnamefont {Wenniger}}, \bibinfo {author} {\bibfnamefont {S.~C.}\ \bibnamefont {Wein}}, \bibinfo {author} {\bibfnamefont {D.}~\bibnamefont {Fioretto}}, \bibinfo {author} {\bibfnamefont {S.~E.}\ \bibnamefont {Thomas}}, \bibinfo {author} {\bibfnamefont {C.}~\bibnamefont {Antón-Solanas}}, \bibinfo {author} {\bibfnamefont {A.}~\bibnamefont {Lemaître}}, \bibinfo {author} {\bibfnamefont {I.}~\bibnamefont {Sagnes}}, \bibinfo {author} {\bibfnamefont {A.}~\bibnamefont {Harouri}}, \bibinfo {author} {\bibfnamefont {N.}~\bibnamefont {Belabas}}, \bibinfo {author} {\bibfnamefont {N.}~\bibnamefont {Somaschi}}, \bibinfo {author} {\bibfnamefont {P.}~\bibnamefont {Hilaire}}, \bibinfo {author} {\bibfnamefont {J.}~\bibnamefont {Senellart}}, \ and\ \bibinfo {author} {\bibfnamefont {P.}~\bibnamefont {Senellart}},\ }\href {\doibase 10.1364/OPTICAQ.527420} {\bibfield  {journal} {\bibinfo  {journal} {Optica Quantum}\ }\textbf {\bibinfo {volume} {2}},\ \bibinfo
  {pages} {404} (\bibinfo {year} {2024})},\ \bibinfo {note} {publisher: Optica Publishing Group}\BibitemShut {NoStop}%
\bibitem [{\citenamefont {Trivedi}\ \emph {et~al.}(2020)\citenamefont {Trivedi}, \citenamefont {Fischer}, \citenamefont {Vu{\v c}kovi{\'c}},\ and\ \citenamefont {M{\"u}ller}}]{trivedi2020generation}%
  \BibitemOpen
  \bibfield  {author} {\bibinfo {author} {\bibfnamefont {R.}~\bibnamefont {Trivedi}}, \bibinfo {author} {\bibfnamefont {K.~A.}\ \bibnamefont {Fischer}}, \bibinfo {author} {\bibfnamefont {J.}~\bibnamefont {Vu{\v c}kovi{\'c}}}, \ and\ \bibinfo {author} {\bibfnamefont {K.}~\bibnamefont {M{\"u}ller}},\ }\href@noop {} {\bibfield  {journal} {\bibinfo  {journal} {Advanced Quantum Technologies}\ }\textbf {\bibinfo {volume} {3}},\ \bibinfo {pages} {1900007} (\bibinfo {year} {2020})}\BibitemShut {NoStop}%
\bibitem [{\citenamefont {Fischer}\ \emph {et~al.}(2017)\citenamefont {Fischer}, \citenamefont {Hanschke}, \citenamefont {Wierzbowski}, \citenamefont {Simmet}, \citenamefont {Dory}, \citenamefont {Finley}, \citenamefont {Vu{\v c}kovi{\'c}},\ and\ \citenamefont {M{\"u}ller}}]{fischer2017signatures}%
  \BibitemOpen
  \bibfield  {author} {\bibinfo {author} {\bibfnamefont {K.~A.}\ \bibnamefont {Fischer}}, \bibinfo {author} {\bibfnamefont {L.}~\bibnamefont {Hanschke}}, \bibinfo {author} {\bibfnamefont {J.}~\bibnamefont {Wierzbowski}}, \bibinfo {author} {\bibfnamefont {T.}~\bibnamefont {Simmet}}, \bibinfo {author} {\bibfnamefont {C.}~\bibnamefont {Dory}}, \bibinfo {author} {\bibfnamefont {J.~J.}\ \bibnamefont {Finley}}, \bibinfo {author} {\bibfnamefont {J.}~\bibnamefont {Vu{\v c}kovi{\'c}}}, \ and\ \bibinfo {author} {\bibfnamefont {K.}~\bibnamefont {M{\"u}ller}},\ }\href@noop {} {\bibfield  {journal} {\bibinfo  {journal} {Nature Physics}\ }\textbf {\bibinfo {volume} {13}},\ \bibinfo {pages} {649} (\bibinfo {year} {2017})}\BibitemShut {NoStop}%
\bibitem [{\citenamefont {Trivedi}\ \emph {et~al.}(2018)\citenamefont {Trivedi}, \citenamefont {Fischer}, \citenamefont {Xu}, \citenamefont {Fan},\ and\ \citenamefont {Vuckovic}}]{trivedi_fewphoton_2018}%
  \BibitemOpen
  \bibfield  {author} {\bibinfo {author} {\bibfnamefont {R.}~\bibnamefont {Trivedi}}, \bibinfo {author} {\bibfnamefont {K.}~\bibnamefont {Fischer}}, \bibinfo {author} {\bibfnamefont {S.}~\bibnamefont {Xu}}, \bibinfo {author} {\bibfnamefont {S.}~\bibnamefont {Fan}}, \ and\ \bibinfo {author} {\bibfnamefont {J.}~\bibnamefont {Vuckovic}},\ }\href {\doibase 10.1103/PhysRevB.98.144112} {\bibfield  {journal} {\bibinfo  {journal} {Physical Review B}\ }\textbf {\bibinfo {volume} {98}},\ \bibinfo {pages} {144112} (\bibinfo {year} {2018})}\BibitemShut {NoStop}%
\bibitem [{\citenamefont {Heuck}\ \emph {et~al.}(2020)\citenamefont {Heuck}, \citenamefont {Jacobs},\ and\ \citenamefont {Englund}}]{heuck_controlledphase_2020}%
  \BibitemOpen
  \bibfield  {author} {\bibinfo {author} {\bibfnamefont {M.}~\bibnamefont {Heuck}}, \bibinfo {author} {\bibfnamefont {K.}~\bibnamefont {Jacobs}}, \ and\ \bibinfo {author} {\bibfnamefont {D.~R.}\ \bibnamefont {Englund}},\ }\href {\doibase 10.1103/PhysRevLett.124.160501} {\bibfield  {journal} {\bibinfo  {journal} {Physical Review Letters}\ }\textbf {\bibinfo {volume} {124}},\ \bibinfo {pages} {160501} (\bibinfo {year} {2020})}\BibitemShut {NoStop}%
\bibitem [{\citenamefont {Das}\ \emph {et~al.}(2019)\citenamefont {Das}, \citenamefont {Zhai}, \citenamefont {{\v C}epulskovskis}, \citenamefont {Javadi}, \citenamefont {Mahmoodian}, \citenamefont {Lodahl},\ and\ \citenamefont {S{\o}rensen}}]{das2019wave}%
  \BibitemOpen
  \bibfield  {author} {\bibinfo {author} {\bibfnamefont {S.}~\bibnamefont {Das}}, \bibinfo {author} {\bibfnamefont {L.}~\bibnamefont {Zhai}}, \bibinfo {author} {\bibfnamefont {M.}~\bibnamefont {{\v C}epulskovskis}}, \bibinfo {author} {\bibfnamefont {A.}~\bibnamefont {Javadi}}, \bibinfo {author} {\bibfnamefont {S.}~\bibnamefont {Mahmoodian}}, \bibinfo {author} {\bibfnamefont {P.}~\bibnamefont {Lodahl}}, \ and\ \bibinfo {author} {\bibfnamefont {A.~S.}\ \bibnamefont {S{\o}rensen}},\ }\href@noop {} {\bibfield  {journal} {\bibinfo  {journal} {arXiv preprint arXiv:1912.08303}\ } (\bibinfo {year} {2019})},\ \Eprint {http://arxiv.org/abs/1912.08303} {arxiv:1912.08303} \BibitemShut {NoStop}%
\bibitem [{\citenamefont {Chang}\ \emph {et~al.}(2007)\citenamefont {Chang}, \citenamefont {S{\o}rensen}, \citenamefont {Demler},\ and\ \citenamefont {Lukin}}]{chang2007single}%
  \BibitemOpen
  \bibfield  {author} {\bibinfo {author} {\bibfnamefont {D.~E.}\ \bibnamefont {Chang}}, \bibinfo {author} {\bibfnamefont {A.~S.}\ \bibnamefont {S{\o}rensen}}, \bibinfo {author} {\bibfnamefont {E.~A.}\ \bibnamefont {Demler}}, \ and\ \bibinfo {author} {\bibfnamefont {M.~D.}\ \bibnamefont {Lukin}},\ }\href@noop {} {\bibfield  {journal} {\bibinfo  {journal} {Nature physics}\ }\textbf {\bibinfo {volume} {3}},\ \bibinfo {pages} {807} (\bibinfo {year} {2007})}\BibitemShut {NoStop}%
\bibitem [{\citenamefont {Shen}\ and\ \citenamefont {Fan}(2007)}]{shen_strongly_2007}%
  \BibitemOpen
  \bibfield  {author} {\bibinfo {author} {\bibfnamefont {J.-T.}\ \bibnamefont {Shen}}\ and\ \bibinfo {author} {\bibfnamefont {S.}~\bibnamefont {Fan}},\ }\href {\doibase 10.1103/PhysRevA.76.062709} {\bibfield  {journal} {\bibinfo  {journal} {Physical Review A}\ }\textbf {\bibinfo {volume} {76}},\ \bibinfo {pages} {062709} (\bibinfo {year} {2007})}\BibitemShut {NoStop}%
\bibitem [{\citenamefont {Witthaut}\ and\ \citenamefont {S{\o}rensen}(2010)}]{witthaut_photon_2010}%
  \BibitemOpen
  \bibfield  {author} {\bibinfo {author} {\bibfnamefont {D.}~\bibnamefont {Witthaut}}\ and\ \bibinfo {author} {\bibfnamefont {A.~S.}\ \bibnamefont {S{\o}rensen}},\ }\href {\doibase 10.1088/1367-2630/12/4/043052} {\bibfield  {journal} {\bibinfo  {journal} {New Journal of Physics}\ }\textbf {\bibinfo {volume} {12}},\ \bibinfo {pages} {043052} (\bibinfo {year} {2010})}\BibitemShut {NoStop}%
\bibitem [{\citenamefont {Mollow}(1975)}]{mollow_purestate_1975}%
  \BibitemOpen
  \bibfield  {author} {\bibinfo {author} {\bibfnamefont {B.~R.}\ \bibnamefont {Mollow}},\ }\href {\doibase 10.1103/PhysRevA.12.1919} {\bibfield  {journal} {\bibinfo  {journal} {Physical Review A}\ }\textbf {\bibinfo {volume} {12}},\ \bibinfo {pages} {1919} (\bibinfo {year} {1975})}\BibitemShut {NoStop}%
\bibitem [{\citenamefont {{\u C}epulkovskis}(2017)}]{mantasthesis}%
  \BibitemOpen
  \bibfield  {author} {\bibinfo {author} {\bibfnamefont {M.}~\bibnamefont {{\u C}epulkovskis}},\ }\href@noop {} {\bibfield  {journal} {\bibinfo  {journal} {Nonlinear photon interactions in waveguides, Master's thesis, Niels Bohr Institute, University of Copenhagen}\ } (\bibinfo {year} {2017})}\BibitemShut {NoStop}%
\bibitem [{\citenamefont {Zhai}(2017)}]{zhaithesis}%
  \BibitemOpen
  \bibfield  {author} {\bibinfo {author} {\bibfnamefont {L.}~\bibnamefont {Zhai}},\ }\href@noop {} {\bibfield  {journal} {\bibinfo  {journal} {Characterisation of Single-Photon sources: fromPulse- Width and Power Dependence of Single-Photon Purity to Effect of Spectral Filtering ,Master's thesis, Niels Bohr Institute, University of Copenhagen}\ } (\bibinfo {year} {2017})}\BibitemShut {NoStop}%
\bibitem [{\citenamefont {Kiraz}\ \emph {et~al.}(2004)\citenamefont {Kiraz}, \citenamefont {Atat{\"u}re},\ and\ \citenamefont {Imamo{\u g}lu}}]{kiraz2004quantum}%
  \BibitemOpen
  \bibfield  {author} {\bibinfo {author} {\bibfnamefont {A.}~\bibnamefont {Kiraz}}, \bibinfo {author} {\bibfnamefont {M.}~\bibnamefont {Atat{\"u}re}}, \ and\ \bibinfo {author} {\bibfnamefont {A.}~\bibnamefont {Imamo{\u g}lu}},\ }\href@noop {} {\bibfield  {journal} {\bibinfo  {journal} {Physical Review A}\ }\textbf {\bibinfo {volume} {69}},\ \bibinfo {pages} {032305} (\bibinfo {year} {2004})}\BibitemShut {NoStop}%
\bibitem [{\citenamefont {Giesz}\ \emph {et~al.}(2016)\citenamefont {Giesz}, \citenamefont {Somaschi}, \citenamefont {Hornecker}, \citenamefont {Grange}, \citenamefont {Reznychenko}, \citenamefont {De~Santis}, \citenamefont {Demory}, \citenamefont {Gomez}, \citenamefont {Sagnes}, \citenamefont {Lemaitre} \emph {et~al.}}]{giesz2016coherent}%
  \BibitemOpen
  \bibfield  {author} {\bibinfo {author} {\bibfnamefont {V.}~\bibnamefont {Giesz}}, \bibinfo {author} {\bibfnamefont {N.}~\bibnamefont {Somaschi}}, \bibinfo {author} {\bibfnamefont {G.}~\bibnamefont {Hornecker}}, \bibinfo {author} {\bibfnamefont {T.}~\bibnamefont {Grange}}, \bibinfo {author} {\bibfnamefont {B.}~\bibnamefont {Reznychenko}}, \bibinfo {author} {\bibfnamefont {L.}~\bibnamefont {De~Santis}}, \bibinfo {author} {\bibfnamefont {J.}~\bibnamefont {Demory}}, \bibinfo {author} {\bibfnamefont {C.}~\bibnamefont {Gomez}}, \bibinfo {author} {\bibfnamefont {I.}~\bibnamefont {Sagnes}}, \bibinfo {author} {\bibfnamefont {A.}~\bibnamefont {Lemaitre}},  \emph {et~al.},\ }\href@noop {} {\bibfield  {journal} {\bibinfo  {journal} {Nature communications}\ }\textbf {\bibinfo {volume} {7}},\ \bibinfo {pages} {1} (\bibinfo {year} {2016})}\BibitemShut {NoStop}%
\bibitem [{\citenamefont {Gardiner}\ and\ \citenamefont {Collett}(1985)}]{gardiner1985input}%
  \BibitemOpen
  \bibfield  {author} {\bibinfo {author} {\bibfnamefont {C.~W.}\ \bibnamefont {Gardiner}}\ and\ \bibinfo {author} {\bibfnamefont {M.~J.}\ \bibnamefont {Collett}},\ }\href@noop {} {\bibfield  {journal} {\bibinfo  {journal} {Physical Review A}\ }\textbf {\bibinfo {volume} {31}},\ \bibinfo {pages} {3761} (\bibinfo {year} {1985})}\BibitemShut {NoStop}%
\bibitem [{\citenamefont {Sekatski}\ \emph {et~al.}(2022)\citenamefont {Sekatski}, \citenamefont {Oudot}, \citenamefont {Caspar}, \citenamefont {Thew},\ and\ \citenamefont {Sangouard}}]{sekatski2022}%
  \BibitemOpen
  \bibfield  {author} {\bibinfo {author} {\bibfnamefont {P.}~\bibnamefont {Sekatski}}, \bibinfo {author} {\bibfnamefont {E.}~\bibnamefont {Oudot}}, \bibinfo {author} {\bibfnamefont {P.}~\bibnamefont {Caspar}}, \bibinfo {author} {\bibfnamefont {R.}~\bibnamefont {Thew}}, \ and\ \bibinfo {author} {\bibfnamefont {N.}~\bibnamefont {Sangouard}},\ }\href {\doibase 10.22331/q-2022-12-13-875} {\bibfield  {journal} {\bibinfo  {journal} {Quantum}\ }\textbf {\bibinfo {volume} {6}},\ \bibinfo {pages} {875} (\bibinfo {year} {2022})}\BibitemShut {NoStop}%
\bibitem [{\citenamefont {Hellebek}\ \emph {et~al.}(2024)\citenamefont {Hellebek}, \citenamefont {M\o{}lmer},\ and\ \citenamefont {S\o{}rensen}}]{hellebek2024}%
  \BibitemOpen
  \bibfield  {author} {\bibinfo {author} {\bibfnamefont {E.~R.}\ \bibnamefont {Hellebek}}, \bibinfo {author} {\bibfnamefont {K.}~\bibnamefont {M\o{}lmer}}, \ and\ \bibinfo {author} {\bibfnamefont {A.~S.}\ \bibnamefont {S\o{}rensen}},\ }\href {\doibase 10.1103/PhysRevA.110.023728} {\bibfield  {journal} {\bibinfo  {journal} {Phys. Rev. A}\ }\textbf {\bibinfo {volume} {110}},\ \bibinfo {pages} {023728} (\bibinfo {year} {2024})}\BibitemShut {NoStop}%
\end{thebibliography}%

\onecolumngrid
\appendix
\newpage
\section{Master equation and equations of motion}\label{app:eq_motion_master_eq}
In this section, we first explicitly derive the stationary state $\hat{\rho}$. From Eqs. \eqref{eq:master_eq} and \eqref{eq:liouvillian} we obtain the following set of differential equations for the density matrix elements
\begin{equation}
  \begin{split}
    \dot{\rho}_{gg} &= \Gamma \rho_{ee} + \frac{i}{2}\left(\Omega^*\rho_{eg}-\Omega\rho_{ge}\right) \\
    \dot{\rho}_{eg} &= -\left( \frac{\Gamma }{2}+\Gamma _d+i \Delta\right)\rho_{eg} - i\frac{\Omega}{2}\left(\rho_{ee}-\rho_{gg}\right) \\
    \dot{\rho}_{ge} &= -\left( \frac{\Gamma}{2}+\Gamma_d-i\Delta\right)\rho_{ge} + i\frac{\Omega^*}{2}\left(\rho_{ee}-\rho_{gg}\right) \\
    \dot{\rho}_{ee} &= -\Gamma \rho_{ee} + i\frac{1}{2}\left(\Omega\rho_{ge}-\Omega^*\rho_{eg}\right)\,.
  \end{split}
\end{equation}
To find the stationary solution, we set $\dot{\hat{\rho}}=0$, and solve the remaining algebraic equations to find
\begin{equation}
  \begin{split}
  \rho_{gg}&=\frac{ (\Gamma +2 \Gamma_d)^2+4 \Delta
  ^2+|\Omega|^2 (1 +2 \Gamma_d/\Gamma)}{(\Gamma +2 \Gamma_d)^2+4 \Delta
  ^2+2 |\Omega|^2 (1 +2 \Gamma_d/\Gamma)} \\
  \rho_{eg}&=\frac{ \Omega (2 \Delta +i (\Gamma +2 \Gamma_d))}{(\Gamma +2 \Gamma_d)^2+4 \Delta
  ^2+2 |\Omega|^2 (1 +2 \Gamma_d/\Gamma)} \\
  \rho_{ge}&=\frac{\Omega^* (2
  \Delta-i(\Gamma +2 \Gamma_d) )}{(\Gamma +2 \Gamma_d)^2+4 \Delta
  ^2+2 |\Omega|^2 (1 +2 \Gamma_d/\Gamma)} \\
  \rho_{ee}&=\frac{|\Omega|^2 (1 +2 \Gamma_d/\Gamma)}{(\Gamma +2 \Gamma_d)^2+4 \Delta ^2+2 |\Omega|^2 (1 +2 \Gamma_d/\Gamma)} \,.
  \label{eq:rho_steady}
  \end{split}
\end{equation}
This stationary solution allows us to calculate the output intensity which we present for $\Gamma_d=0$ in the main text \eqref{eq:intensity}. For completeness, we here include the solution for non-zero pure dephasing rate $\Gamma_d$
\begin{equation}
 \frac{I_{\text{out}}}{I_{\text{in}}} = A\left[ \frac{4}{(1+2\Gamma_d/\Gamma)^2+\left(\frac{2\Delta}{\Gamma}\right)^2 + 2\frac{I}{I_{\text{sat}}}(1+2\Gamma_d/\Gamma)}\left(1+2\Gamma_d/\Gamma - (|x|e^{-i\theta}\left(\frac{\Gamma+2\Gamma_d}{2\Gamma}+i\frac{\Delta}{\Gamma}\right)+\text{c.c.})\right)+\abs{x}^2\right] \,.
 \label{eq:intensity_app}
\end{equation}

Next we detail how Eqs. \eqref{eq:eq_motion} are obtained from Schrödinger's equation. By inserting the wavefunction ansatz \eqref{eq:ansatz} into the Schrödinger equation we can derive relatively simple equations for the wavefunction's components
\begin{align}
\begin{split}
  \dot{c}_g(t) &= i\frac{\Omega}{2}^*c_e(t) \\
  \dot{c}_e(t) &= -i\Delta c_e(t)+i\frac{\Omega}{2}c_g(t)+i\sqrt{\Gamma}\phi_{1,g}(t,t) \\
  \dot{\phi}_{1,g}(t,t_{e1}) &= i\sqrt{\Gamma}\delta(t-t_{e1}) c_e(t)+i\frac{\Omega}{2}^*\phi_{1,e}(t,t_{e1}) \\
  \dot{\phi}_{1,e}(t,t_{e1}) &= -i\Delta\phi_{1,e}(t,t) + i\frac{\Omega}{2}\phi_{1,g}(t,t_{e1})+i\sqrt{\Gamma}\phi_{2}(t,t,t_{e1}) \\
  \dot{\phi}_{2}(t,t_{e2},t_{e1}) &= i\sqrt{\Gamma}\phi_{1,e}(t,t_{e1})\delta(t-t_{e2})\,.
  \label{eq:eq_motion_app}
  \end{split}
\end{align}
Recall that we have introduced the decay rate $\Gamma= 2\pi \mathcal{G}^2/v_g$ and the Rabi frequency $\Omega= 2\sqrt{2\pi}\mathcal{E}\mathcal{G}$. These equations of motion are the same as the well known equations of a driven two-level system with a decaying excited state. For the initial condition we assume the emitter to be in the ground state $\ket{ \psi'(t=0)}=\ket{g,\varnothing}$ corresponding to $c_g(t=0)=1$. The equations of motion are solved numerically are in a rotated frame with respect to those obtained in Eq.~\eqref{eq:eq_motion_app} in order to have the faster numerical convergence. In particular, we introduce the rotated functions $\tilde{c}_e(t) = \exp(i\Delta t)c_e(t)$ and $\tilde{\phi}_{1,e}(t,t_{e,1}) = \exp(i\Delta t)\phi_{1,e}(t,t_{e,1})$. After integrating $\dot{\phi}_{1,g}(t,t_{e,1})$ and $\dot{\phi}_{2}(t,t_{e,2},t_{e,1})$ in the time domain before ($0<t<t_{e,1}-\epsilon$) and after ($t_{e,1}+\epsilon<t<\infty$) the first photon emission respectively, we obtain Eqs. \eqref{eq:eq_motion}.

\section{Full expressions of the correlation functions}\label{app:G1_G2}
Here we include the full expressions of the unnormalised first and second order correlation functions $G^{(1)}(t,t_1,t_2)$ and $G^{(2)}(t,t_1,t_2)$ for the wavefunction ansatz expressed in Eq.~\eqref{eq:ansatz} by means of Eqs. \eqref{eq:G1} and \eqref{eq:G2}:
\begin{equation}
\begin{split}
  G^{(1)}(t,t_1,t_2) &= \bra{ \psi'(t)}\left( \hat{E}^{\dagger}(t_1) + \mathcal{E}^{*}(t_1) \right)\left( \hat{E}(t_2) + \mathcal{E}(t_2) \right) \ket{ \psi'(t)} \\
  &= \phi^{*}_{1,g}(t,t-t_{1})\phi_{1,g}(t,t-t_{2}) + \mathcal{E}^{*}(t_1)\mathcal{E}(t_2)\\
  &+ \int dt_e \bm{\Big(}\phi^{*}_{2}(t,t_e,t-t_{1}) + \phi^{*}_{2}(t,t-t_{1},t_e)\bm{\Big)}\bm{\Big(}\phi_{2}(t,t_e,t-t_{2}) + \phi_{2}(t,t-t_{2},t_e)\bm{\Big)}\\
  &+ \mathcal{E}^{*}(t_1)\Bigg( c_g^{*}(t)\phi_{1,g}(t,t-t_{2}) + \int dt_e \phi^{*}_{1,g}(t,t_e)\bm{\Big(}\phi_{2}(t,t_e,t-t_{2}) + \phi_{2}(t,t-t_{2},t_e)\bm{\Big)}\Bigg)\\
  &+ \mathcal{E}(t_2)\Bigg( c_g(t)\phi^{*}_{1,g}(t,t-t_{1}) + \int dt_e \phi_{1,g}(t,t_e)\bm{\Big(}\phi^{*}_{2}(t,t_e,t-t_{1}) + \phi^{*}_{2}(t,t-t_{1},t_e)\bm{\Big)}\Bigg), \\
  \quad \\
  G^{(2)}(t,t_1,t_2) &= \bra{ \psi'(t)}\left( \hat{E}^{\dagger}(t_1) + \mathcal{E}^{*}(t_1) \right)\left( \hat{E}^{\dagger}(t_2) + \mathcal{E}^{*}(t_2) \right)\left( \hat{E}(t_2) + \mathcal{E}(t_2) \right)\left( \hat{E}(t_1) + \mathcal{E}(t_1) \right) \ket{ \psi'(t)} \\
  &= |\phi_{2}(t,t-t_{2},t-t_{1})|^2 + |\phi_{2}(t,t-t_{1},t-t_{2})|^2 + |\mathcal{E}(t_1)|^2 |\mathcal{E}(t_2)|^2\\
  &+ |\mathcal{E}(t_1)|^2\int dt_e \bm{\Big(}|\phi_{2}(t,t_e,t-t_{2})|^2 + |\phi_{2}(t,t-t_{2},t_e)|^2 + (\phi^{*}_{2}(t,t-t_{2},t_e)\phi_{2}(t,t_e,t-t_{2}) + H.c.) \bm{\Big)}\\
  &+ |\mathcal{E}(t_2)|^2\int dt_e \bm{\Big(}|\phi_{2}(t,t_e,t-t_{1})|^2 + |\phi_{2}(t,t-t_{1},t_e)|^2 + (\phi^{*}_{2}(t,t-t_{1},t_e)\phi_{2}(t,t_e,t-t_{1}) + H.c.) \bm{\Big)}\\
  &+ |\mathcal{E}(t_1)|^2|\phi_{1,g}(t,t-t_{2})|^2 + |\mathcal{E}(t_2)|^2|\phi_{1,g}(t,t-t_{1})|^2 \\
  &+ \Bigg( \phi^{*}_{2}(t,t-t_{2},t-t_{1})\phi_{2}(t,t-t_{1},t-t_{2}) + \mathcal{E}(t_1)\mathcal{E}(t_2)c_g(t)\bm{\Big(}\phi^{*}_{2}(t,t-t_{1},t-t_{2})\ + \phi^{*}_{2}(t,t-t_{2},t-t_{1}) \bm{\Big)} \\
  &+ \bm{\Big(}\mathcal{E}(t_1)\phi^{*}_{1,g}(t,t-t_{2}) + \mathcal{E}(t_2)\phi^{*}_{1,g}(t,t-t_{1})\bm{\Big)}\bm{\Big(}\phi^{*}_{2}(t,t-t_{2},t-t_{1})+\phi^{*}_{2}(t,t-t_{2},t-t_{1})\bm{\Big)} \\
  &+ \mathcal{E}(t_1)\mathcal{E}^{*}(t_2)\bm{\Big(} 
  \phi^{*}_{1,g}(t,t-t_{1})\phi_{1,g}(t,t-t_{2}) + \int dt_e ( \phi^{*}_{2}(t,t_e,t-t_{1})\phi_{2}(t,t_e,t-t_{2}) \\
  &+ \phi^{*}_{2}(t,t-t_{1},t_e)\phi_{2}(t,t_e,t-t_{2}) + \phi^{*}_{2}(t,t_e,t-t_1)\phi_{2}(t,t-t_{2},t_e) + \phi^{*}_{2}(t,t-t_{1},t_e)\phi_{2}(t,t-t_{2},t_e) ) \bm{\Big)} \\
  &+ |\mathcal{E}(t_1)|^2\mathcal{E}(t_2)\int dt_e \phi_{1,g}(t,t_e)\bm{\Big(}\phi^{*}_{2}(t,t_e,t-t_{2}) + \phi^{*}_{2}(t,t-t_{2},t_e) \bm{\Big)}\\
  &+ |\mathcal{E}(t_2)|^2\mathcal{E}(t_1)\int dt_e \phi_{1,g}(t,t_e)\bm{\Big(}\phi^{*}_{2}(t,t_e,t-t_{1}) + \phi^{*}_{2}(t,t-t_{1},t_e) \bm{\Big)}\\
  &+ |\mathcal{E}(t_1)|^2\mathcal{E}(t_2) c_g(t) \phi^{*}_{1,g}(t,t-t_{2}) + |\mathcal{E}(t_2)|^2\mathcal{E}(t_1) c_g(t) \phi^{*}_{1,g}(t,t-t_{1}) + H.c. \Bigg)\,.
\end{split}
\end{equation}

\section{Filtering \label{app:filtering}}

In this Appendix we describe in more detail how we filter the output field in our analysis. The transformation where we apply the transmission function $\mathcal{T}(\omega-\omega_c)$ is done in the Heisenberg picture, as shown in the main text through Eq.~\eqref{eq:Heis_filter}. The correlation functions in the frequency domain thus satisfy
\begin{equation}\begin{split}
  G_F^{(1)}(\omega_1,\omega_2)&=\langle \hat{E}^{(f)\dagger}_{out}(\omega_1) \hat{E}^{(f)}_{out}(\omega_2) \rangle 
  = T^*(\omega_1-w_c)T(\omega_2-w_c) \langle \hat{E}_{out}^\dagger(\omega_1)\hat{E}_{out}(\omega_2) \rangle, \\[6pt]
  G_F^{(2)}(\omega_1,\omega_2,\omega_3,\omega_4)&=\langle \hat{E}^{(f)\dagger}_{out}(\omega_1) \hat{E}^{(f)\dagger}_{out}(\omega_2) \hat{E}^{(f)}_{out}(\omega_3) \hat{E}^{(f)}_{out}(\omega_4) \rangle \\
  &= T^*(\omega_1-w_c)T^*(\omega_2-w_c)T(\omega_3-w_c)T(\omega_4-w_c) \langle \hat{E}_{out}^\dagger(\omega_1)\hat{E}_{out}^\dagger(\omega_2)\hat{E}_{out}(\omega_3)\hat{E}_{out}(\omega_4) \rangle\,,
  \label{eq:G1_G2_F}
\end{split}
\end{equation}
where the index ($F$) stands for \textit{filtered}. In our numerical simulation, however, we calculate the correlation functions instead through Eqs.~\eqref{eq:G1} and \eqref{eq:G2}, which depend on the one- and two-photon wavefunctions $\phi_{1,g}(t,t_1)$ and $\phi_2(t,t_1,t_2)$ as well as the classical field $\mathcal{E}(t)$. Therefore we use that the transformation in Eq.~\eqref{eq:Heis_filter} for the wavefunctions translates into
\begin{align}
\begin{split}
    \phi^{(f)}_{1,g}(\omega,\omega_1) &= \mathcal{T}(\omega_1-\omega_c)\phi_{1,g}(\omega,\omega_1),\\
    \phi^{(f)}_{2}(\omega,\omega_1,\omega_2) &= \mathcal{T}(\omega_1-\omega_c)\mathcal{T}(\omega_2-\omega_c)\phi_{2}(\omega,\omega_1,\omega_2),\\
    \mathcal{E}^{(f)}(\omega) &= \mathcal{T}(\omega-\omega_c)\mathcal{E}(\omega)\,,
  \label{eq:T_wavefunctions}
\end{split}
\end{align}
where we have previously calculated
\begin{equation}
     \phi_{1,g}(\omega,\omega_1)=\mathcal{F}\{\phi_{1,g}(t,t_1)\}, \quad 
    \phi_{2}(\omega,\omega_1,\omega_2)=\mathcal{F}\{\phi_{2}(t,t_1,t_2)\}, \quad
    \mathcal{E}(\omega) = \mathcal{F}\{\mathcal{E}(t)\}\,,
\end{equation}
with $\mathcal{F}$ indicating the 1- or 2-D Fourier transform. The correlation functions are then obtained by applying the inverse Fourier transform over Eqs.\eqref{eq:T_wavefunctions} and inserting $\phi^{(f)}_{1,g}(t,t_1)$, $\phi^{(f)}_{2}(t,t_1,t_2)$ and $\mathcal{E}^{(f)}(t)$ into Eqs.~\eqref{eq:G1} and \eqref{eq:G2} accordingly.

In addition, it must also be noted from Eqs. \eqref{eq:G1_G2_F} that there are also terms involving
\begin{align}
  \begin{split}
    M_1(t_1,t_2)&\equiv\int dt_e \big( 
    \phi^{*}_2(t_e,t_1) + \phi^{*}_2(t_1,t_e)
    \big)\big( 
    \phi_2(t_e,t_2) + \phi_2(t_2,t_e)
    \big), \\
    M_2(t_1,t_2)&\equiv\int dt_e 
    \phi_{1,g}(t_e)
    \big( 
    \phi_2^{*}(t_e,t_1) + \phi_2^{*}(t_1,t_e)
    \big)\mathcal{E}(t_2)\,, \\
  \end{split}
\end{align}
that we define as $M_{1,2}(t_1,t_2)$. These terms correspond to two and one photon contributions from the emitter respectively, but from crossed contributions from the one and two-photon wavefunctions $\phi_{1,g}(t_1)$ and $\phi_{2}(t_1,t_2)$ and thus are filtered according to
\begin{align}
\begin{split}
    M_{1,2}^{(f)}(\omega_1,\omega_2) &= \mathcal{T}^{*}(\omega_1-\omega_c)\mathcal{T}(\omega_2-\omega_c)M_{1,2}(\omega_1,\omega_2)\,,
  \label{eq:T_M_functions}
\end{split}
\end{align}
where we have similarly calculated $M_{1,2}(\omega_1,\omega_2)=\mathcal{F}\{M_{1,2}(t_1,t_2)\}$.

\section{Bounds on the relation between HOM visibility and multiphoton emission}\label{app:flimits_reducedV}

To evaluate the limits of $F$, we first rewrite the visibility in Eq.\eqref{eq:final_VHOM} to
\begin{equation}
  \begin{split}
    V &=1-  \frac{\iint dt_e dt_e' {\left( G^{(2)}(t_e,t_e')+G^{(1)}(t_e,t_e)G^{(1)}(t_e',t_e')- \abs{G^{(1)}(t_e,t_e')}^2\right)}}{\iint dt_e dt_e' \left( G^{(2)}(t_e,t_e')+G^{(1)}(t_e,t_e)G^{(1)}(t_e',t_e')\right)}\\
    &=1-\frac{1}{(1+g^{(2)})}{\left[g^{(2)} +\frac{\iint dt_e dt_e' {\left(G^{(1)}(t_e,t_e)G^{(1)}(t_e',t_e')- \abs{G^{(1)}(t_e,t_e')}^2\right)}}{\langle n\rangle  ^{2}}\right]}
    \,.
  \end{split}
  \label{eq:VHOM_1minus}
\end{equation}
where $\langle n\rangle$ is the mean number of photons in the pulse. Here we see that the HOM visibility is reduced directly from the two photon component through $g^{(2)}$ but also by any deviation from perfect coherence, i.e. if $G^{(1)}(t_e,t_e)G^{(1)}(t_e',t_e')> \abs{G^{(1)}(t_e,t_e')}^2$. 

We  split the density matrix for the outgoing light field into three components  $\hat{\rho} = {P_0} \hat{\rho}_0 + P_1 \hat{\rho}_1 + P_{n>1} \hat{\rho}_{n>1}$. Here $\hat{\rho}_0  $,  $ \hat{\rho}_1$, and $ 
\hat{\rho}_{n>1}$ correspond to (normalised) density matrices containing zero, one and more than one photon respectively. $P_i$ with $i=1$, $2$ and $>$ denote the corresponding probabilities and fulfil $\sum_i P_i =1$. We note that this form ignores any coherence between different number states, but such coherences are irrelevant in the current context, since all expectation values conserve the number of photons. 

The density matrix for the single photon component is Hermitian and  can hence be diagonalised into a complete set of modes
\begin{align}
\begin{split}
\hat{\rho}_1 &= \sum_l  w_l\iint dt_1dt_2  \psi_l^{*}(t_1)\psi_l(t_2)\hat{a}^{\dagger}(t_1)\ket{\emptyset}\bra{\emptyset}\hat{a}(t_2) \,,
\end{split}
\end{align}
where $\psi_l(t)$ are the (orthonormal) single photon mode functions and their weights $w_l$ fulfil $\sum_l w_l=1$. With this form of the density matrix, we can evaluate the first order coherence function for the single photon state
\begin{equation}
    G^{(1)}_1(t_e,t_e')=\langle \hat{a}^\dag(t_e)\hat{a}(t_e')\rangle_1=\sum_l w_l \psi_l(t_e)\psi_l^*(t_e')   \,,
\end{equation}
where subscripts $i$ refer to expectation value with respect to $\hat\rho_i$. 

We now substitute the decomposition of the density matrix into the integral in the last line of Eq.~\eqref{eq:VHOM_1minus} to obtain
\begin{equation}
    \begin{split}
            \iint dt_e dt_e'&{\left(    G^{(1)}(t_e,t_e)G^{(1)}(t_e',t_e')- \abs{G^{(1)}(t_e,t_e')}^2\right)}=\\
            &P_1^2 (1-V_1)             +2 P_1P_{n>1}{\left(\langle \hat n\rangle_{n>1}- \sum_l w_l \iint dt_e dt_e'  \psi_l(t_e)\psi_l^*(t_e') \langle a^\dagger(t_e)\hat a(t_e')\rangle_{n>1} \right)}+ P_{n>1}^2 I_{n>1},    \end{split}
\label{eq:VHOMintegral}
\end{equation}
where $V_1=\sum_l w_l^2$ is visibility of the single photon component, $\hat n=\int dt \hat a^\dagger(t)\hat a(t)$ is the photon number operator,  and 
\begin{equation}
    I_{n>1}= \iint dt_e dt_e'{\left(    G_{n>1}^{(1)}(t_e,t_e)G_{n>1}^{(1)}(t_e',t_e')- \abs{G_{n>1}^{(1)}(t_e,t_e')}^2\right)},
\end{equation}
express the incoherence of the multiphoton state, which is not so important for the discussion below since we assume that we have a good single photon source such that $P_{n>1}\ll 1$. 
Introducing the single mode operators
\begin{equation}
    \hat a_l=\int dt \psi_l^*(t)\hat a(t),
\end{equation}
the above expression can be simplified to 
 \begin{equation}
    \begin{split}
             \iint dt_e dt_e'{\left(    G^{(1)}(t_e,t_e)G^{(1)}(t_e',t_e')- \abs{G^{(1)}(t_e,t_e')}^2\right)}=P_1^2(1-V_1)+2 P_1P_{n>1}\sum_l w_l {\left(\langle \hat n\rangle_{n>1}-\langle \hat n_l \rangle_{n>1} \right)}+P_{n>1}^2 I_{n>1},
    \end{split}
\label{eq:VHOMintegral_final}
    \end{equation}
where $\hat n_l =\hat a_l^\dagger \hat a_l$ is the number operator for the mode $\psi_l$. From this expression, we see that the reduction in visibility from the multi-photon component is caused by the number of photons in other modes than the single photon state $\langle \hat n_{\neq l}\rangle_{n>1}=\langle \hat n\rangle_{n>1}-\langle \hat n_l \rangle_{n>1}$. Substituting back into Eq.~\eqref{eq:VHOM_1minus} we obtain
\begin{equation}
  \begin{split}
    V  
    =1-\frac{1}{1+g^{(2)}}[g^{(2)}
    +\frac{P_1^2(1-V_1)+2P_1P_{n>1} \sum_l w_l\langle \hat n_{\neq l}\rangle_{n>1} +P_{n>1}^2 I_{n>1}}{\langle \hat n\rangle  ^{2}}]
    \,.
  \end{split}
  \label{eq:VHOM_1minus_simplified}
\end{equation}
To relate  the above result to $g^{(2)}$, we note that 
\begin{equation}
    g^{(2)}=\frac{P_{n>1} \langle \hat n(\hat n-1)\rangle_{n>1}}{\langle \hat n\rangle^2}.
\end{equation}
We can thus eliminate $P_{n>1}$ from Eq.~\eqref{eq:VHOM_1minus_simplified} to obtain
\begin{equation}
  \begin{split}
    V  
    =V_1(1-F{g^{(2)}})+B(1-V_1)g^{(2)}
    \, ,
  \end{split}
  \label{eq:VHOM_F_full}
\end{equation}
where 
\begin{equation}
  \begin{split}
    F=1 +P_1 \frac{2 \sum_l w_l\langle \hat n_{\neq l}\rangle_{n>1}}{\langle \hat n(\hat n-1)\rangle_{n>1}}\,,
  \end{split}
  \label{eq:Ffull}
\end{equation}
and 
\begin{equation}
     B=2P_1\sum_l w_1 \frac{\langle \hat n_l\rangle_{n>1}}{\langle \hat n(\hat n-1)\rangle_{n>1}}\,.
\end{equation}
Here, we have for simplicity only kept the leading order terms in $g^{(2)}$. 

For typical single photon sources, $\hat \rho_{n>1}$ will be dominated by the two-photon component $n=2$. In this case, the expression for $F$ in Eq.\eqref{eq:Ffull} reduces to the expression in Eq.~\eqref{eq:F_fully_simplified} if we consider a pure state ($w_1=1$ and $w_l$=0, for $l\neq 1)$ and replace the subscript by $\neq \psi$ for the particular mode function $\psi=\psi_1$ that we consider. We note, however, that since  $n(n-1)\geq n$ for $n\geq 2$, we have  $0\leq \langle \hat n_{\neq l}\rangle_{n>1}\leq \langle \hat n\rangle_{n>1} \leq \langle \hat n(\hat n-1)\rangle_{n>1}$ and the value of $F$ is always in the interval $F\in [1,3]$, regardless of these simplifications.  

We also note that we could equivalently have expressed Eq.~\eqref{eq:VHOM_F_full} in the form $V=V_1-F g^{(2)}+B'(1-V_1)g^{(2)} $ with the same $F$. In this case we would have 
\begin{equation}
    B'=1+2P_1\frac{\langle  \hat n\rangle_{n>1} }{\langle \hat n( \hat n-1)\rangle_{n>1}}.
\end{equation}
From here we see that if $\hat \rho_{n>1}$ is dominated by the two-photon component, we have $B'\approx 3$ for $P_1\approx 1$, whereas $B$ is in the interval $B\in [0,2]$. Since $B<B'$ the small cross term $\sim(1-V_1)g^{(2)}$ that we ignore, is smaller for the form we use in Eq.~\eqref{eq:F_factor}. This is consistent with Ref. \cite{hellebek2024}, where it was noted that this form gave a better separation of different physical effects for single photon sources based on spontaneous parametric down conversion. In that work, a value $F<1$ is calculated, which can be attributed to a sizable deviation of $V_1$ from unity, such that the term $B(1-V_1)g^{(2)}$ cannot be neglected.  

\section{Filtering using the quantum regression theorem}
In this section, we model the effects of a Lorentzian filter in the quantum regression theorem by coupling the quantum dot output field to a cavity. 
Although limited to only Lorentzian filtering due to the cavity profile, it provides a useful way to check the accuracy of the wavefunction ansatz. 

To model the cavity and quantum dot system, we use the standard Pauli spin operators, acting on a $2\times 2$ Hilbert space. We restrict the cavity creation and annihilation operators $a^\dag$ and $a$ to a maximum of two excitations, i.e., a $3\times 3$ Hilbert space when including zero excitation.
The resulting density matrix $\rho $
is composed of the density matrix for the quantum dot and the cavity. 
The Hamiltonian that describes the resulting dynamics is $H = H_\mathrm{QD} + H_\mathrm{cav} + H_\mathrm{int}$, where 
\begin{align}
\begin{split}
    H_{QD} &= \frac{\Omega}{2} \left(\mathbf{I}_3 \otimes \sigma_x \right)  + \Delta \left( \mathbf{I}_3 \otimes \sigma_{ee} \right)
    \\ H_\mathrm{cav} &= \omega_\mathrm{cav} \left( a^\dag a \otimes \mathbf{I}_2\right)
    \\ H_\mathrm{int} &= \frac{\sqrt{\Gamma \kappa}}{2 \sqrt{2}} \left( a ^{\dagger} \otimes \sigma_- + a \otimes \sigma_+\right),
    \end{split}
\end{align}
where $\mathbf{I}_n$ is the $n \times n$ identity matrix. 
We then define the Liouville operators (where the tensor products are suppressed for brevity)
\begin{align}
    c_1 &= i \sqrt{\Gamma} \sigma_- + \sqrt{\frac{\kappa}{2}} a
    \\c_2 &=\sqrt{\frac{\kappa}{2}}a,
\end{align}
with the corresponding master equation in Lindblad form
\begin{align}
\dot \rho = -i \left[H,\rho \right]+ \sum_k \left( c_k \rho c_k^{\dag} - \frac{1}{2} \left\{ c_k^{\dag}c_k, \rho \right\} \right).
\label{eq: master_equation_lindblad}
\end{align}
Our definitions yield an effective Hamiltonian 
\begin{align}
H_\mathrm{eff} &= H_\mathrm{int} - \sum_{k} \frac{i}{2}c_k^{\dag}c_k
\\ &= -\frac{i\Gamma}{2} \sigma_+\sigma_- -\frac{i \kappa}{2} a^{\dag} a + \sqrt{\frac{\kappa \Gamma}{2}} a^{\dag}\sigma_-.
\end{align}
We include the leaked field via the output operator
\begin{align}
    a_{out} &= \sqrt{\frac{\kappa}{2}}a + x \int_0^t \frac{\kappa}{2} \mathcal{E}(t') e^{-(\kappa/2+i \omega_{cav})(t-t')} dt',
\end{align}
where $\mathcal{E}(t) = \frac{\Omega(t)}{2\sqrt{\Gamma v_g}}$ is defined as in the main text. 
Inserting the form of $\Omega(t)$ as in~\eqref{eq: omega_pulse} and assuming we are on resonance with the cavity, i.e., $\omega_{cav} =0$, we find

\begin{align}
a_\mathrm{out} = \sqrt{\frac{\kappa}{2}} a	+ x \frac{\pi  \kappa  e^{\frac{1}{8} \kappa  \left(\kappa  \sigma ^2-4 t+4 t_0\right)} \left(\text{erf}\left(\frac{\kappa  \sigma ^2+2 t_0}{2 \sqrt{2} \sigma }\right)-\text{erf}\left(\frac{\kappa  \sigma ^2-2 t+2 t_0}{2 \sqrt{2} \sigma }\right)\right)}{8 \sqrt{\Gamma  v_g}},
\end{align}
where $a$ is the non-leaked cavity field modelled by the master equation.
We will use the form 
\begin{align}
    a_\mathrm{out} &= \sqrt{\frac{\kappa}{2}}a + x \alpha(t),
\end{align}
in the following. 
Below, we write down the terms that are evaluated with the quantum regression theorem, where we solve \eqref{eq: master_equation_lindblad} to find the time-dependence of the density matrix $\rho(t)$.

\subsection{$G^{(2)}(t_1,t_2)$ terms}
Evaluating $G^{(2)}(t_1,t_2)$ with the output operator $a_\mathrm{out}$ gives several terms, which we write collected by powers of the leaked field $x$. 

\begin{align}
\begin{split}
\text{I.  }  &\frac{\kappa^2}{4} \text{Tr}\left( a(t_2) a(t_1) \rho a^{\dag}(t_1) a^{\dag}(t_2) \right)\\
\text{II.  } &\frac{\kappa^{3/2}\alpha(t_1) x^{*}}{2 \sqrt{2}} \text{Tr} \left( a(t_2) a(t_1) \rho a^{\dag}(t_2) \right)  + H.c\\
& \frac{\kappa^{3/2}\alpha(t_2) x^{*}}{2 \sqrt{2}} \text{Tr} \left( a(t_2) a(t_1) \rho a^{\dag}(t_1) \right) + H.c.
\\
\text{III.  } &\frac{\kappa}{2}|x|^2 \alpha^2(t_1) \text{Tr}\left( a(t_2) \rho a^{\dag}(t_2)\right) \\
&\frac{\kappa}{2}|x|^2 \alpha^2(t_2) \text{Tr}\left( a(t_1) \rho a^{\dag}(t_1)\right) \\
& \frac{\kappa}{2}|x|^2 \alpha(t_1) \alpha(t_2) \text{Tr} \left( a(t_2) \rho a^{\dag}(t_1) \right) + H.c.\\
\text{IV.   } & \frac{\kappa}{2} (x^{*})^2 \alpha(t_1) \alpha(t_2) \text{Tr}\left(a(t_2) a(t_1) \rho) \right) +H.c \\
\text{V.  } &\frac{\sqrt{\kappa}}{2} x(x^{*})^2 \alpha(t_2)a^2(t_1) \text{Tr}\left(a(t_2) \rho \right) +H.c.\\
&\frac{\sqrt{\kappa}}{2} x(x^{*})^2 \alpha(t_1)\alpha^2(t_2) \text{Tr}\left(a(t_1) \rho \right) +H.c.\\
\text{VI.   } & |x|^4 \alpha^{2}(t_1) \alpha^2(t_2)
\end{split}
\end{align}

\subsection{$G^{(1)}(t_1,t_2)$ and $|G^{(1)}(t_1,t_2)|^2$}

Evaluating $G^{(1)}(t_1,t_2)$ with the output operators yields, in powers of $x$
\begin{align}
\begin{split}
\text{I.   } &\frac{\kappa}{2} \text{Tr}\left( a(t_1) \rho a^{\dag}(t_2)\right) 
\\ \text{II.   } & x \frac{\sqrt{\kappa}}{\sqrt{2}} \alpha(t_1) \text{Tr} \left( \rho a^{\dag}(t_{2})\right) +H.c.
\\ \text{III.   } & |x|^{2} \alpha(t_1) \alpha(t_2).
\end{split}
\end{align}

To compute the visibility, we require the squared modulus 
\begin{align}
\begin{split}
|G^{(1)}(t_1,t_2)|^2 &= \left| \text{Tr} \left( a(t_1) \rho a^\dag(t_2) \right) \right|^2
\\ &= \text{Tr} \left( a(t_2) \rho a^\dag(t_1) \right)\text{Tr} \left( a(t_1) \rho a^\dag(t_2) \right) 
\end{split}
\end{align}

We now use $\text{Tr}(a+b) = \text{Tr}(a) + \text{Tr}(b)$ to split the trace into terms with respect to $x$. 
We find

\begin{align}
\begin{split}
&\text{I.} \, \, \,  \frac{\kappa^2}{4} \text{Tr}\left(a(t_2) \rho a(t_2)^{\dagger }\right) \text{Tr}\left(a(t_1)  \rho  a(t_2)^{\dagger }\right)
\\ &\text{II.    }  \frac{\kappa ^{3/2} x^*}{2\sqrt{2}} \left(\alpha (t_1)\text{Tr}\left( a(t_2) \rho \right) \text{Tr}\left( a(t_1)  \rho  a(t_2)^{\dagger }\right) + \alpha (t_2)\text{Tr}\left( a(t_1) \rho \right) \text{Tr}\left(a(t_2)  \rho  a(t_1)^{\dagger }\right)\right) + H.c. 
\\ &\text{III.   } \frac{\kappa}{2}|x|^2 \alpha(t_1) \alpha(t_2) \left( \left|\text{Tr} \left(\rho a^{\dag}(t_2) \right)\right|^{2} + \left| \text{Tr} \left(\rho a^{\dag}(t_1) \right)\right|^{2} + 2 \text{Tr} \left(a(t_1) \rho a^{\dag}(t_2) \right) \right)  
\\ & \text{IV.    } \frac{\kappa}{2} (x^{*})^2 \alpha(t_1) \alpha(t_2) \text{Tr}\left( a(t_1) \rho\right) \text{Tr}\left( a(t_2) \rho \right) + H.c.
\\ &\text{V.    } \sqrt{\frac{\kappa}{2}} x (x^{*})^2 \left( \alpha^2(t_1) \alpha(t_2) \text{Tr} \left( a(t_2) \rho \right)  + \alpha(t_1) \alpha^2(t_2) \text{Tr} \left( a(t_1) \rho \right)  \right) + H.c. 
\\ &\text{VI.    } |x|^4 \alpha^2(t_1) \alpha^2(t_2)
\end{split}
\end{align}

\end{document}